\documentclass[review]{elsarticle}

\usepackage{hyperref}
\usepackage{siunitx}
\usepackage{booktabs}
\usepackage{xcolor}
\usepackage[normalem]{ulem}
\newcommand{\Kirill}[1]{\colorbox{yellow}{#1}}

\journal{Journal of Magnetism and Magnetic Materials}

\begin{document}

\begin{frontmatter}

\title{Efficient calculation of the mutual inductance of arbitrarily oriented circular filaments via a generalisation of the Kalantarov-Zeitlin method }


\author[imt]{Kirill V. Poletkin\corref{cor1}}
\ead{kirill.poletkin@kit.edu}
\cortext[cor1]{Corresponding author}

\author[imt]{ Jan G. Korvink}

\address[imt]{The Institute of Microstructure Technology, Karlsruhe Institute of
Technology, Hermann-von-Helmholtz-Platz 1, 76344 Eggenstein-Leopoldshafen, Germany}


\begin{abstract}
In this article, we present a new analytical formulation for calculation of the mutual inductance between two circular filaments arbitrarily oriented with respect to each other,  as an alternative to Grover  \cite{Grover1944} and Babi\v{c} \cite{BabicSiroisAkyelEtAl2010} expressions reported in 1944 and 2010, respectively. The formula is derived via a generalisation of the Kalantarov-Zeitlin method, which showed that the calculation of mutual inductance between a circular primary filament and any other secondary filament having an arbitrary shape and any desired position with respect to the primary filament is reduced to a line integral. In particular, the obtained formula provides a solution for the singularity issue arising in the Grover and Babi\v{c} formulas
for the case when the planes of the primary and secondary circular filaments are mutually perpendicular. The efficiency
and flexibility of the Kalantarov-Zeitlin method allow us to extend immediately the application of the obtained result to
a case of the calculation of the mutual inductance between a primary circular filament and its projection on a tilted plane. Newly developed formulas have been successfully  validated through a number of examples available in the literature, and by a direct comparison with the results of calculation performed by the \textit{FastHenry}  software.
\end{abstract}


\end{frontmatter}


\section{Introduction}

Analytical and semi-analytical methods in the calculation of {inductances}, and in particular the mutual inductances of filament wires and their loops, play an important role in power transfer, wireless communication, and sensing and actuation, and is applied in different fields of science, including electrical and electronic engineering, medicine, physics, nuclear magnetic resonance, mechatronics and robotics, to name the most prominent. Collections of formulas for the calculation of mutual inductance between filaments of different geometrical shapes covering a wide spectrum of practical arrangements have variously been presented in classical handbooks by Rosa \cite{Rosa1908}, Grover \cite{Grover2004}, Dwight \cite{Dwight1945}, Snow \cite{Snow1954}, Zeitlin \cite{Zeitlin1950}, Kalantarov \cite{Kalantarov1986}, among others.

The availability of efficient numerical methods such as  \textit{FastHenry} \cite{KamonTsukWhite1994} (based on the multipole expansion) currently provides an accurate and fast solution for the calculation of mutual and self-inductance for any circumstance, including the use of arbitrary materials, conductor cross-sections, loop shapes, and arrangements. However, analytical methods allow to obtain the result in the form of a final formula with a finite number of input parameters, which when applicable may significantly reduce computation effort. It will also facilitate mathematical analysis, for example when derivatives of the mutual inductance w.r.t.\ one or more parameters are required to evaluate electromagnetic forces via the stored magnetic energy, or when optimization is performed.

Analytical methods applied to the calculation of the mutual inductance between two circular filaments is a prime example, and has been successfully used in an increasing number of applications, including electromagnetic levitation \cite{OkressWroughtonComenetzEtAl1952}, superconducting levitation \Kirill{\cite{Urman1997,Urman1997a,Coffey2001}}, \Kirill{magnetic force interaction \cite{Urman2014}}, wireless power transfer \cite{JowGhovanloo2007,SuLiuHui2009,ChuAvestruz2017}, electromagnetic actuation \cite{ShiriShoulaie2009,RavaudLemarquandLemarquand2009,Obata2013,ShalatiPoletkinKorvinkEtAl2018}, micro-machined contactless inductive suspensions \cite{Poletkin2013,Poletkin2014a,Lu2014,PoletkinLuWallrabeEtAl2017b} and hybrid suspensions \cite{Poletkin2012,PoletkinShalatiKorvinkEtAl2018,PoletkinKorvink2018}, biomedical applications \cite{TheodoulidisDitchburn2007,SawanHashemiSehilEtAl2009}, topology optimization \cite{KuznetsovGuest2017}, nuclear magnetic resonance \cite{D.I.B.2002,SpenglerWhileMeissnerEtAl2017}, indoor positioning systems \cite{AngelisPaskuAngelisEtAl2015}, navigation sensors \cite{WuJeonMoonEtAl2016}, and magneto-inductive wireless communications \cite{Gulbahar2017}.

The original formula of the mutual inductance between two coaxial circular filaments was derived by Maxwell \cite[page 340, Art. 701]{Maxwell1954} and expressed in terms of elliptic integrals. Butterworth obtained a formula covering the case of circular filaments with parallel axes \cite{ButterworthM.Sc.1916}. Then, a general formal expression made for cases where the axes of the circles are parallel, and where their axes intersect, was derived by Snow \cite{Snow1928}. However, the Butterworth and Snow expressions suffer from a low rate of convergence. This issue was recognized and solved by Grover, who developed the most general method in the form of a single integral \cite{Grover1944}. Using the vector potential method, as opposed to the Grover means, the general case for calculating the mutual inductance between inclined circular filaments arbitrarily positioned with respect to each other was subsequently obtained by Babi\v{c} et al. \cite{BabicSiroisAkyelEtAl2010}.

Kalantarov and Zeitlin showed that the calculation of mutual inductance between a circular primary filament and any other secondary filament having an arbitrary shape and any desired position with respect to the primary filament can be reduced to a line integral \cite[Sec. 1-12, page 49]{Kalantarov1986}. In the present paper, we report an adaptation of this method to the case of two circular filaments and then derive a new analytical formula for calculating the mutual inductance between two circular filaments having any desired position with respect to each other as an alternative to the Grover and Babi\v{c} expressions.

\begin{figure}[!t]
  \centering
  \includegraphics[width=2.7in]{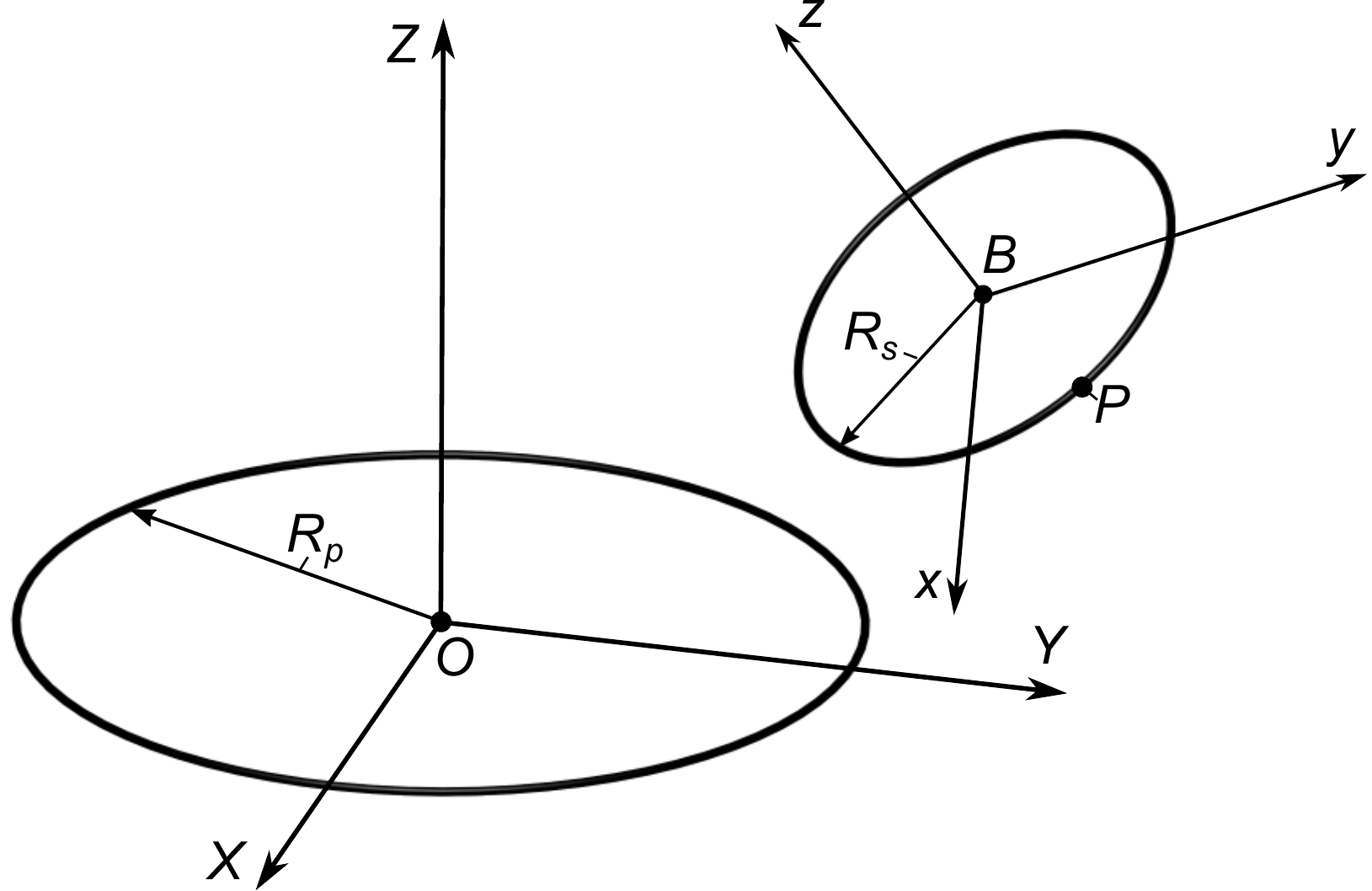}
  \caption{General scheme of arbitrarily positioning two circular filaments with respect to each other: $P$ is an arbitrary point on the secondary filament.   }\label{fig:scheme}
\end{figure}

In particular, the obtained formula provides a solution for the singularity issue arising in the Grover and Babi\v{c} formulas for the case when the planes of the primary and secondary circular filaments are mutually perpendicular. The efficiency and flexibility of the Kalantarov-Zeitlin method allow us to extend immediately the application of the obtained result to a case of the calculation of the mutual inductance between a primary circular filament and its projection on a tilted plane. For instance, this particular case appears in micro-machined inductive suspensions and has a direct practical application in studying their stability \cite{PoletkinLuWallrabeEtAl2017b} and pull-in dynamics \cite{PoletkinShalatiKorvinkEtAl2018,PoletkinKorvink2018}. The new analytical formulae were verified by comparison with series of reference examples covering all cases given by Grover\cite{Grover2004}, Kalantarov and Zeitlin \cite{Kalantarov1986}, and using  direct numerical calculations performed by the Babi\v{c} Matlab function \cite{BabicSiroisAkyelEtAl2010} and the \textit{FastHenry} software \cite{KamonTsukWhite1994}.

\section{Preliminary discussion}
\label{sec:Pleliminary}

Two circular filaments having radii of $R_p$ and $R_s$ for the primary circular filament (the primary circle) and the secondary circular filament (the secondary circle), respectively are considered to be arbitrarily positioned in space, namely, they have a linear and angular misalignment,  as is shown in Figure \ref{fig:scheme}. Let us assign a coordinate frame (CF) denoted as $XYZ$ to the primary circle in a such way that the $Z$ axis is coincident with the circle axis and the $XOY$ plane of the CF lies on the circle's plane, where the origin $O$ corresponds to the centre of primary circle. In turn, the $xyz$ CF is assigned to the secondary circle in a similar way so that its origin $B$ is coincident with the centre of the secondary circle.

\begin{figure}[!t]
  \centering
  \includegraphics[width=3.3in]{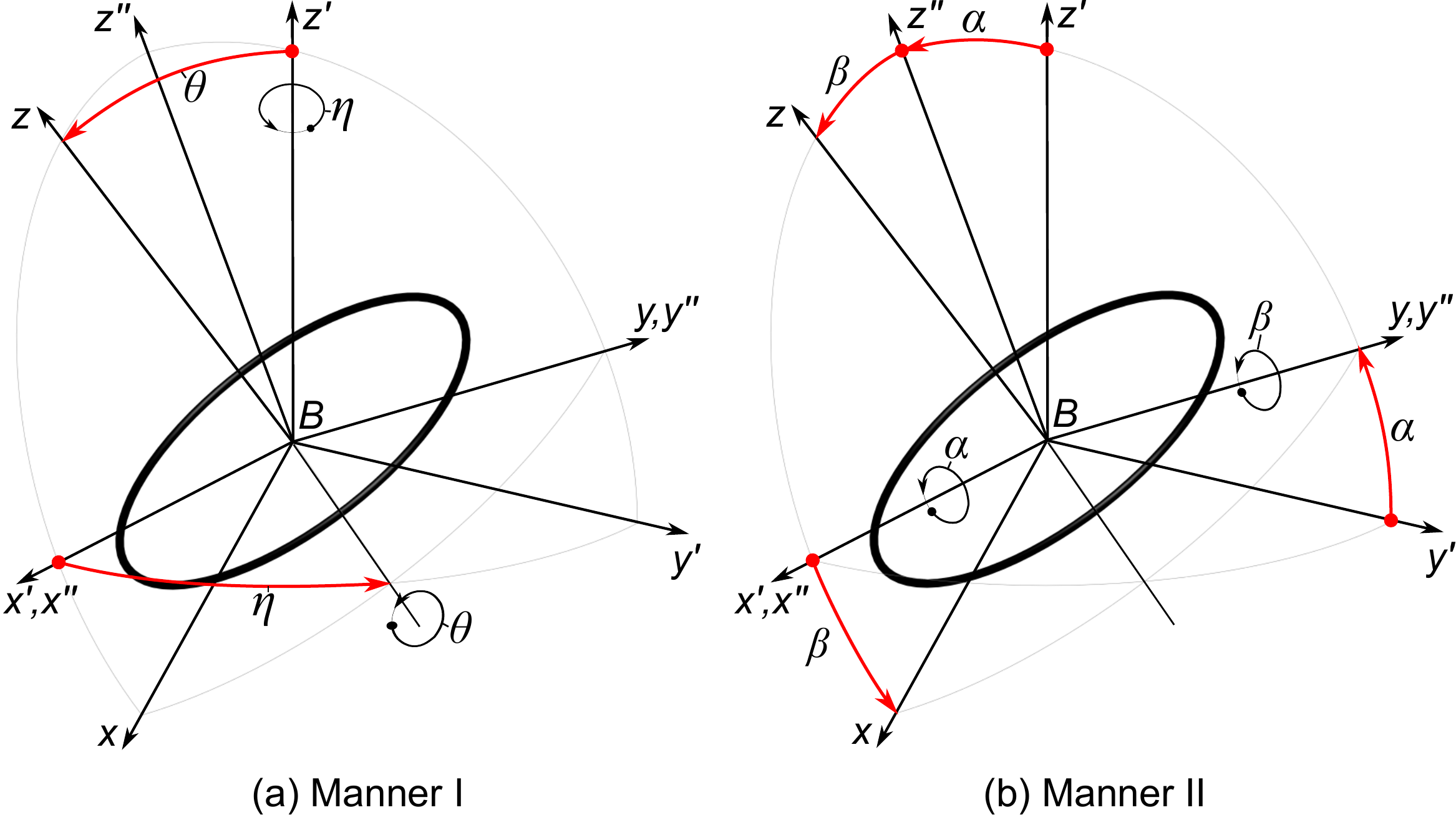}
  \caption{Two manners for determining the angular position of the secondary circle with respect to the primary one: $x'y'z'$ is the auxiliary CF the axes of which are parallel to the axes of $XYZ$, respectively; $x''y''z''$ is the auxiliary CF defined in such a way that the $x'$ and $x''$ are coincide, but the $z''$ and $y''$ axis is rotated by the $\alpha$ angle with respect to the $z'$ and $y'$ axis, respectively.   }\label{fig:angular position}
\end{figure}

The linear position of the secondary circle with respect to the primary one is defined by the coordinates of the centre $B$ ($x_B,y_B,z_B$). The angular position of the secondary circle can be defined in two ways. Firstly, the angular position is defined by the angle $\theta$  and $\eta$ corresponding to the angular rotation around an axis passing through the diameter of the secondary circle, and then the rotation of this axis lying on the surface $x'By'$ around the vertical $z'$ axis, respectively, as it is shown in Figure \ref{fig:angular position}(a). These angles for determination of angular position of the secondary circle was proposed by Grover and used in his formula numbered by (179) in \cite[page 207]{Grover2004} addressing the general case for calculation of the mutual inductance between two circular filaments.

The same angular position can be determined through the $\alpha$ and $\beta$ angle, which corresponds to the angular rotation around the $x'$ axis and then around the $y''$ axis, respectively, as it is shown in Figure \ref{fig:angular   position}(b). This additional second  manner  is more convenient in a case of study dynamics and stability issues, for instance, applying to axially symmetric inductive levitation systems \cite{Poletkin2014a,PoletkinLuWallrabeEtAl2017b} in compared with   the Grover manner.  These two pairs of angles have the following relationship with respect to each other such as:
\begin{equation}\label{eq:angles}
  \left\{\begin{array}{l}
   \sin\beta=\sin\eta\sin\theta;\\
   \cos\beta\sin\alpha=\cos\eta\sin\theta.
  \end{array}\right.
\end{equation}

The details of the derivation of this set presented above are shown in   \ref{app:determination}.

\begin{figure}[!t]
  \centering
  \includegraphics[width=2.8in]{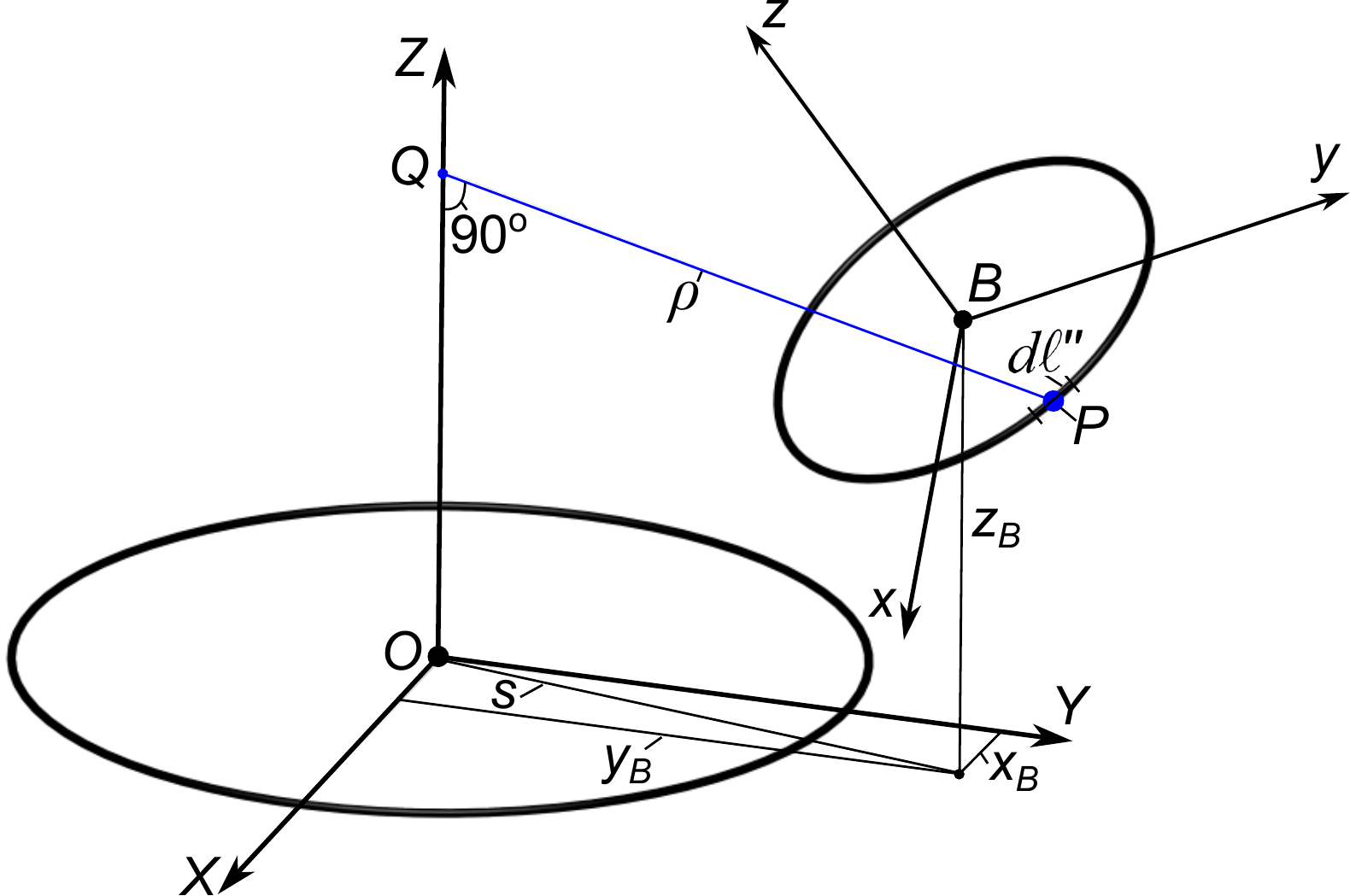}
  \caption{The Kalantarov-Zeitlin method: $s=\sqrt{x_B^2+y_B^2}$ is the distance to the centre $B$ on the $XOY$ plane.   }\label{fig:KZ_method}
\end{figure}

\section{The Kalantarov-Zeitlin method}

Using the general scheme for two circular filaments  shown in Figure \ref{fig:scheme} as an illustrative one, the Kalantarov-Zeitlin method is presented.
The method reduces  the calculation of mutual inductance between a circular primary filament and any other secondary filament having an arbitrary shape and any desired position with respect to the primary circular filament to a line integral \cite[Sec. 1-12, page 49]{Kalantarov1986}.

\begin{figure}[!t]
  \centering
  \includegraphics[width=1.8in]{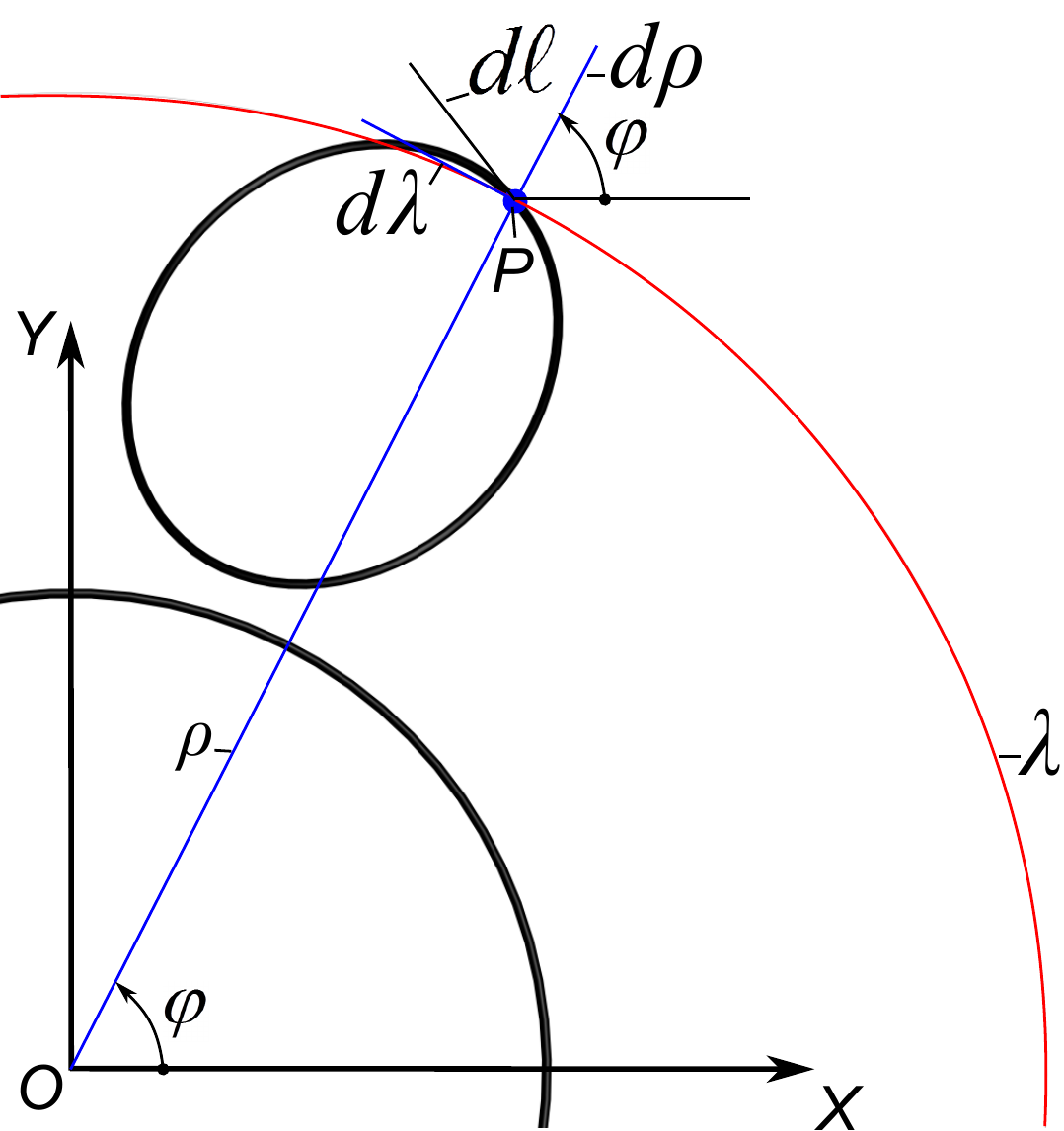}
  \caption{The Kalantarov-Zeitlin method: projection of the secondary filament on the $\rho$-plane passed through the point $P$ and parallel to the plane of the primary circular filament; $d\ell$ is the projection of the element $d\ell''$ on the $\rho$-plane.   }\label{fig:KZ_method plane}
\end{figure}

Indeed, let us choose an arbitrary point $P$ of the secondary filament (as it has been mentioned above the filament can have any shape), as shown in Figure \ref{fig:scheme}. An element of length $d\ell''$ of the secondary filament at the point $P$ is considered.
Also, the point $P$ is connected the point $Q$ lying on the $Z$ axis by a line, which is perpendicular to the $Z$ axis and has a length of $\rho$, as shown in Figure \ref{fig:KZ_method}. Then the element $d\ell''$ can be decomposed on $dz$ along the $Z$ axis and on $d\rho$ along the $\rho$ line and $d\lambda$ along the $\lambda$-circle having radius of $\rho$ (see, Figure \ref{fig:KZ_method plane}). It is  obvious that the mutual inductance between $dz$ and the primary circular filament is equal to zero because $dz$ is perpendicular to a plane of primary circle. But the mutual inductance between $d\rho$ and the primary circular filament is also equal to zero because of the symmetry of the  primary circle relative to the $d\rho$ direction.

Thus, the  mutual inductance $dM$ between element $d\ell''$  and the primary circle is equal to the mutual inductance $dM_{\lambda}$ between element $d\lambda$ and the primary circle. Moreover, due to the fact that the primary and the $\lambda$-circle are coaxial and, consequently, symmetric then we can write:
\begin{equation}\label{eq:KZ_method}
  \frac{dM_{\lambda}}{M_{\lambda}}=\frac{d\lambda}{\lambda}=\frac{d\lambda}{2\pi\rho},
\end{equation}
where $M_\lambda$ is the mutual inductance of the primary \Kirill{coil} and $\lambda$-circle.

From Figure \ref{fig:KZ_method plane}, it is directly seen that
\begin{equation}\label{eq:KZ_lambda}
  d\lambda=dy\cos\varphi-dx\sin\varphi=(\cos\zeta\cos\varphi-\cos\varepsilon\sin\varphi)d\ell,
\end{equation}
where $\cos\varepsilon$ and $\cos\zeta$ are the direction cosines of element $d\ell$ relative to the $X$ and $Y$ axis, respectively.  Hence, accounting for (\ref{eq:KZ_method}) and (\ref{eq:KZ_lambda}), we can write:
\begin{equation}\label{eq:dM}
  dM=dM_{\lambda}=M_{\lambda}\frac{\cos\zeta\cos\varphi-\cos\varepsilon\sin\varphi}{2\pi\rho}d\ell,
\end{equation}
and as a result, a line integral for calculation mutual inductance between the primary circle and a filament is
\begin{equation}\label{eq:M}
  M=\frac{1}{2\pi}\int_{\ell}M_{\lambda}\frac{\cos\zeta\cos\varphi-\cos\varepsilon\sin\varphi}{\rho}d\ell,
\end{equation}
 where $M_\lambda$  is defined by the Maxwell formula for mutual inductance between two coaxial circles \cite[page 340, Art. 701]{Maxwell1954}. Note that during integrating,  the $Z$ coordinate  of the element $d\ell$ is also changing and this dependency is taken into account by the $M_\lambda$ function directly.


\begin{figure}[!t]
  \centering
  \includegraphics[width=2.5in]{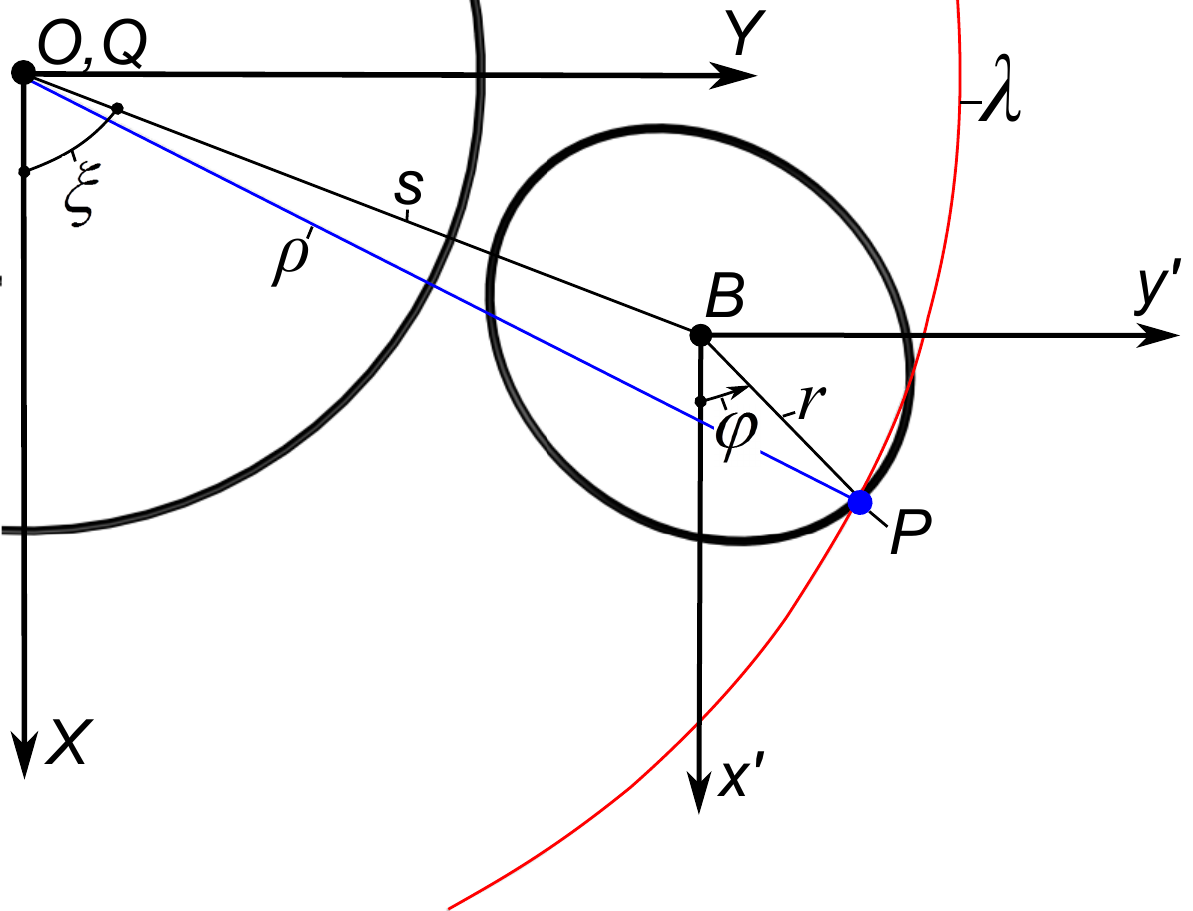}
  \caption{Determination of the position of the point $P$ on the $\rho$-plane through the fixed parameter $s$ and the distance $r$.   }\label{fig:derivation plane}
\end{figure}

\begin{figure}[!b]
  \centering
  \includegraphics[width=2.0in]{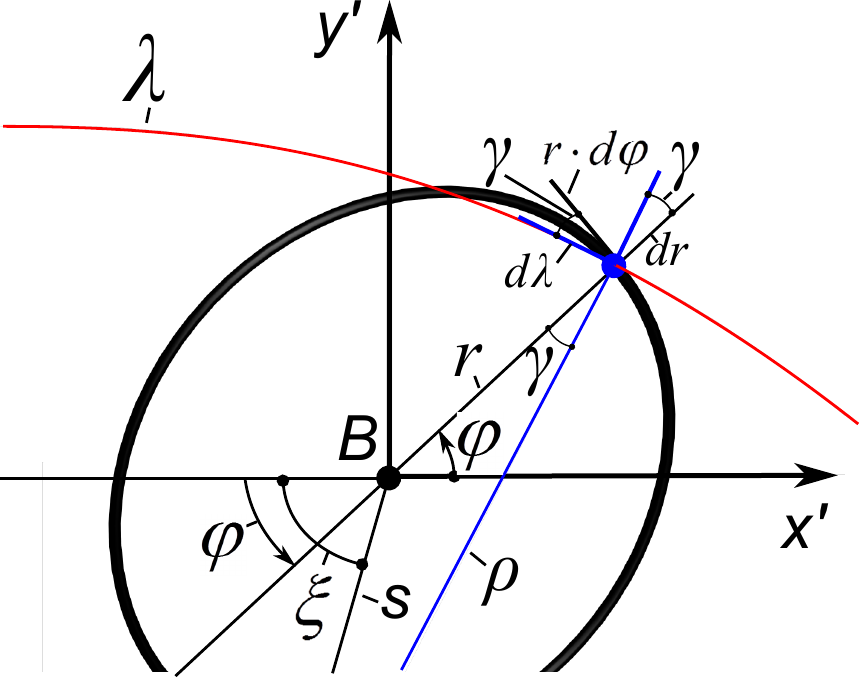}
  \caption{The relationship between $d\lambda$ and $d\varphi$.   }\label{fig:lambda and phi}
\end{figure}

%
%

\section{Derivation of Formulas}
Due to the particular geometry of secondary filament under consideration, its projection on the $\rho$-plane (the $\rho$-plane is parallel to the primary circle plane and passed through the point $P$) is an ellipse, which can be defined in a polar coordinate by a function $r=r(\varphi)$ with the origin at the point $B$ as it is shown in Figure \ref{fig:derivation plane}. Hence, the distance $\rho$ can be expressed in terms of the parameter $s$, which is fixed, and the distance $r$ from the origin $B$, which is varied with the angular variable $\varphi$. Introducing the angle $\gamma$ as shown in Figure \ref{fig:lambda and phi}, for the distance $\rho$ the following equations can be written:
\begin{equation}\label{eq:rho cos}
  \begin{array}{l}
     \rho \cos\gamma=r+s\cos(\xi-\varphi), \\
    \rho \sin\gamma=s\sin(\xi-\varphi).
  \end{array}
\end{equation}
Due to (\ref{eq:rho cos}), we have:
\begin{equation}\label{eq:rho sq}
      \rho^2 =r^2+r\cdot s\cos(\xi-\varphi)+s^2,
\end{equation}
where the function $r=r(\varphi)$ can be defined as \cite{SpiegelLipschutzLiu2009}:
\begin{equation}\label{eq:r}
     r=\frac{R_s\cos\theta}{\sqrt{\sin^2(\varphi-\eta)+\cos^2\theta\cos^2(\varphi-\eta)}}.
\end{equation}
The angle $\theta$ and $\eta$ defines the angular position of the secondary circle with respect to the primary one according to manner I considered in Sec. \ref{sec:Pleliminary}. Note that the function $r$ can be also defined via the angles  $\alpha$ and $\beta$ of manner II  also considered in Sec. \ref{sec:Pleliminary} as  it is shown in  \ref{app:formulas}. However, for the further derivation, the angular position of the secondary circle is defined through manner I, since it is convenient for the direct comparison with Grover's and Babi\v{c}' results.

According to Figure \ref{fig:lambda and phi}, the relationship between the element $d\lambda$ of the $\lambda$-circle and an increment  of the angle $\varphi$ is as follows:
\begin{equation}\label{eq:d_lambda and d_vi}
  d\lambda=r\cdot d\varphi\cos\gamma-dr\sin\gamma=\left(r\cos\gamma-\frac{dr}{d\varphi}\sin\gamma\right)d\varphi.
\end{equation}

Then, accounting for (\ref{eq:d_lambda and d_vi}), (\ref{eq:rho sq}) and (\ref{eq:rho cos}), line integral (\ref{eq:M}) can be replaced by a definite integral for the calculation of mutual inductance as follows:
\begin{equation}\label{eq:def integral}
  M=\frac{1}{2\pi}\int_{0}^{2\pi}M_{\lambda}\frac{r^2+r\cdot s\cos(\xi-\varphi)-\frac{dr}{d\varphi}s\sin(\xi-\varphi)}{\rho^2}d\varphi.
\end{equation}

\begin{figure}[!t]
  \centering
  \includegraphics[width=2.0in]{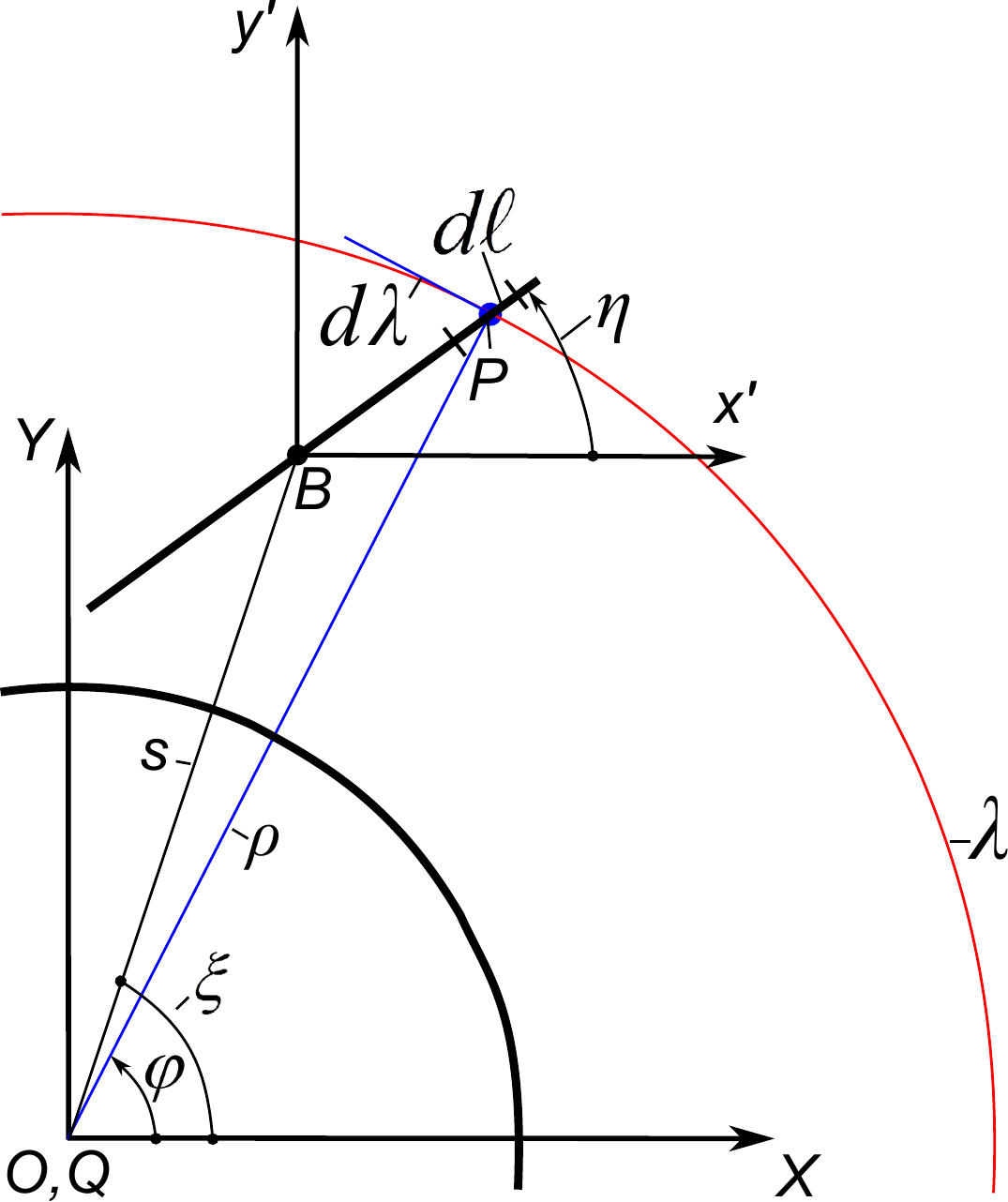}
  \caption{The special case: the two filament circles are mutually perpendicular to each other.   }\label{fig:special case}
\end{figure}

Now, let us introduce the following dimensionless parameters such as:
\begin{equation}\label{eq:dimensionless_par}
  \begin{array}{l}
   {\displaystyle \bar{x}_B=\frac{x_B}{R_s};\; \bar{y}_B=\frac{y_B}{R_s};\; \bar{z}_B=\frac{z_B}{R_s};\;\bar{r}=\frac{r}{R_s};} \\
   {\displaystyle  \bar{\rho}=\frac{\rho}{R_s};\bar{s}=\sqrt{\bar{x}_B^2+\bar{y}_B^2}.}
  \end{array}
\end{equation}
The $\varphi$-derivative of $\bar{r}$ is
\begin{equation}\label{eq:derivative}
  \frac{d\bar{r}}{d\varphi}=\frac{1}{2}\bar{r}^3\tan^2\theta\sin(2(\varphi-\eta)),
\end{equation}
The mutual inductance $M_{\lambda}$ is
\begin{equation}\label{eq:M_lambda}
M_{\lambda}=\mu_0\frac{2}{k} \Psi(k)\sqrt{R_pR_s\bar{\rho}},
\end{equation}
where  $\mu_0$ is the magnetic permeability of free space, and
\begin{equation}\label{eq:Maxell}
   \Psi(k)=\left(1-\frac{k^2}{2}\right)K(k)-E(k),
\end{equation}
where $K(k)$ and $E(k)$ are the complete elliptic functions of the first and second kind, respectively,
and
\begin{equation}\label{eq:k}
   k^2=\frac{4\nu\bar{\rho}}{(\nu\bar{\rho}+1)^2+\nu^2\bar{z}_{\lambda}^2},
\end{equation}
where $\nu=R_s/R_p$ and $\bar{z}_{\lambda}=\bar{z}_B+\bar{r}\tan\theta\sin(\varphi-\eta)$. Accounting for dimensionless parameters (\ref{eq:dimensionless_par}) and substituting (\ref{eq:derivative}) and (\ref{eq:Maxell}) into integral (\ref{eq:def integral}),
the new formula to calculate the mutual inductance between two circular filaments having any desired position with respect to each other becomes
\begin{equation}\label{eq:NEW FORMULA}
  M=\frac{\mu_0\sqrt{R_pR_s}}{\pi}\int_{0}^{2\pi}\frac{\bar{r}+t_1\cdot\cos\varphi+t_2\cdot\sin\varphi}{k\bar{\rho}^{1.5}}\cdot\bar{r}\cdot\Psi(k)d\varphi,
\end{equation}
where terms $t_1$ and $t_2$ are defined as
\begin{equation}\label{eq:t}
  \begin{array}{l}
    t_1=\bar{x}_B+0.5\bar{r}^2\tan^2\theta\sin(2(\varphi-\eta))\cdot\bar{y}_B; \\
     t_2=\bar{y}_B-0.5\bar{r}^2\tan^2\theta\sin(2(\varphi-\eta))\cdot\bar{x}_B,
  \end{array}
\end{equation}
and  $\bar{\rho} =\sqrt{\bar{r}^2+2\bar{r}\cdot \bar{s}\cos(\xi-\varphi)+\bar{s}^2}$.

Formula (\ref{eq:NEW FORMULA}) can be applied to any possible cases, but one is excluded when the two filament circles are mutually perpendicular to each other. In this case the projection of the secondary circle onto the $\rho$-plane becomes simply a line as it is shown in Fig. \ref{fig:special case} and as a result to integrate with respect to $\varphi$ is no longer possible.

For the treatment of this  case, the   Kalantarov-Zeitlin formula (\ref{eq:M}) is directly used. Let us introduce the dimensionless variable  $\bar{\ell}=\ell/R_s$ and then the   integration of (\ref{eq:M}) is preformed with respect to this dimensionless variable $\bar{\ell}$
within interval  $-1\leq\bar{\ell}\leq1$.
The direction cosines $\cos\zeta$  and $\cos\varepsilon$ become as $\sin\eta$  and $\cos\eta$, respectively (see, Fig. \ref{fig:special case}).
Accounting for
\begin{equation}\label{eq: varphi}
  \begin{array}{l}
     \rho \cos\varphi=s\cos\xi+\ell\cos\eta, \\
    \rho \sin\varphi=s\sin\xi +\ell\sin\eta,
  \end{array}
\end{equation}
and the Maxwell formula (\ref{eq:Maxell}) and (\ref{eq:k}), where the $Z$-coordinate of the element $d\bar{\ell}$ is defined as
\begin{equation}\label{eq: Z}
  \bar{z}_{\lambda}=\bar{z}_B\pm\sqrt{1-\bar{\ell}^2},
\end{equation}
then the formula to calculate the mutual inductance between two   filament circles, which are mutually perpendicular to each other,  becomes
as follows:
\begin{equation}\label{eq:Singular case}
 \begin{array}{l}
  {\displaystyle M=\frac{\mu_0\sqrt{R_pR_s}}{\pi}\left[\int_{-1}^{1}\frac{t_1-t_2}{k\bar{\rho}^{1.5}}\cdot\Psi(k)d\bar{\ell}\right.}\\
  {\displaystyle  \left.+\int_{1}^{-1}\frac{t_1-t_2}{k\bar{\rho}^{1.5}}\cdot\Psi(k)d\bar{\ell}\right]},
  \end{array}
\end{equation}
where terms $t_1$ and $t_2$ are defined as
\begin{equation}\label{eq:t for singular case}
  \begin{array}{l}
    t_1=\sin\eta(\bar{x}_B+\bar{\ell}\cos\eta); \\
     t_2=\cos\eta(\bar{y}_B+\bar{\ell}\sin\eta),
  \end{array}
\end{equation}
and  $\bar{\rho} =\sqrt{\bar{s}^2+2\bar{\ell}\cdot \bar{s}\cos(\xi-\eta)+\bar{\ell}^2}$. Note that integrating (\ref{eq:Singular case}) between $-1$ and $1$ equation (\ref{eq: Z}) is calculated with the positive sign and for the other direction the negative sign is taken.

In order to demonstrate the efficiency and flexibility of the    Kalantarov-Zeitlin method, a formula for the calculation of the mutual inductance between the primary circular filament and its projection on a tilted plane is obtained as follows. In this case, the function of $r=r(\varphi)$ is constant and defined through the radius of primary coil as  $r=R_p$. Since the centre of the projection is coincide with the $Z$-axis, thus $s=0$. Then, the formula is derived from (\ref{eq:NEW FORMULA}) as its particular case ($\bar{s}=0$ and $\bar{r}=\bar{\rho}=1$) and  becomes,  simply,
 \begin{equation}\label{eq:PROJECTOIN FORMULA}
  M=\frac{\mu_0{R_p}}{\pi}\int_{0}^{2\pi}\frac{1}{k}\cdot\Psi(k)d\varphi.
\end{equation}
The obtained formulas can be easily programmed, they are intuitively understandable for application. Also, the singularity arises in Grover's and Babi\v{c}'s formula for the calculate of the mutual inductance between two   filament circles, which are mutually perpendicular to each other,  is solved in developed formula (\ref{eq:Singular case}).
The \textit{Matlab} files with the implemented formulas (\ref{eq:NEW FORMULA}), (\ref{eq:Singular case}) and (\ref{eq:PROJECTOIN FORMULA}) are available from the authors  as   supplementary materials to this article. Also, in  \ref{app:formulas} the  the developed formulas can be rewritten through the pair of the angle $\alpha$ and $\beta$.

\section{ Examples of Calculation. Numerical Verification }
In this section,  developed new formulas (\ref{eq:NEW FORMULA}), (\ref{eq:Singular case}) and (\ref{eq:PROJECTOIN FORMULA}) are verified by the examples taken from  Grover \cite{Grover2004} and  Kalantarov \cite{Kalantarov1986} books and Babi\v{c} article  \cite{BabicSiroisAkyelEtAl2010}. The special attention was addressed to the singularity case arisen when the two filament circles are mutually perpendicular to each other. Then, formula  (\ref{eq:PROJECTOIN FORMULA}) was validated with the \textit{FastHenry} software \cite{KamonTsukWhite1994}.
All  calculations for considered cases proved the robustness and efficiency of developed formulas.

Note that  the notation  proposed in Grover's and Kalantarov's books in order to define the linear misalignment of the secondary coil is different from the notation used in the Babi\v{c} article and in our article as well.
Also,  the angular misalignment in the Babi\v{c} formula must be defined through the parameters of the secondary coil plane.
These particularities of the notation will be discussed specifically for each case.  For all calculation, the primary coil is located on the plane $XOY$ and its centre at the origin $O$(0,0,0).

\begin{figure}[!t]
  \centering
  \includegraphics[width=1.45in]{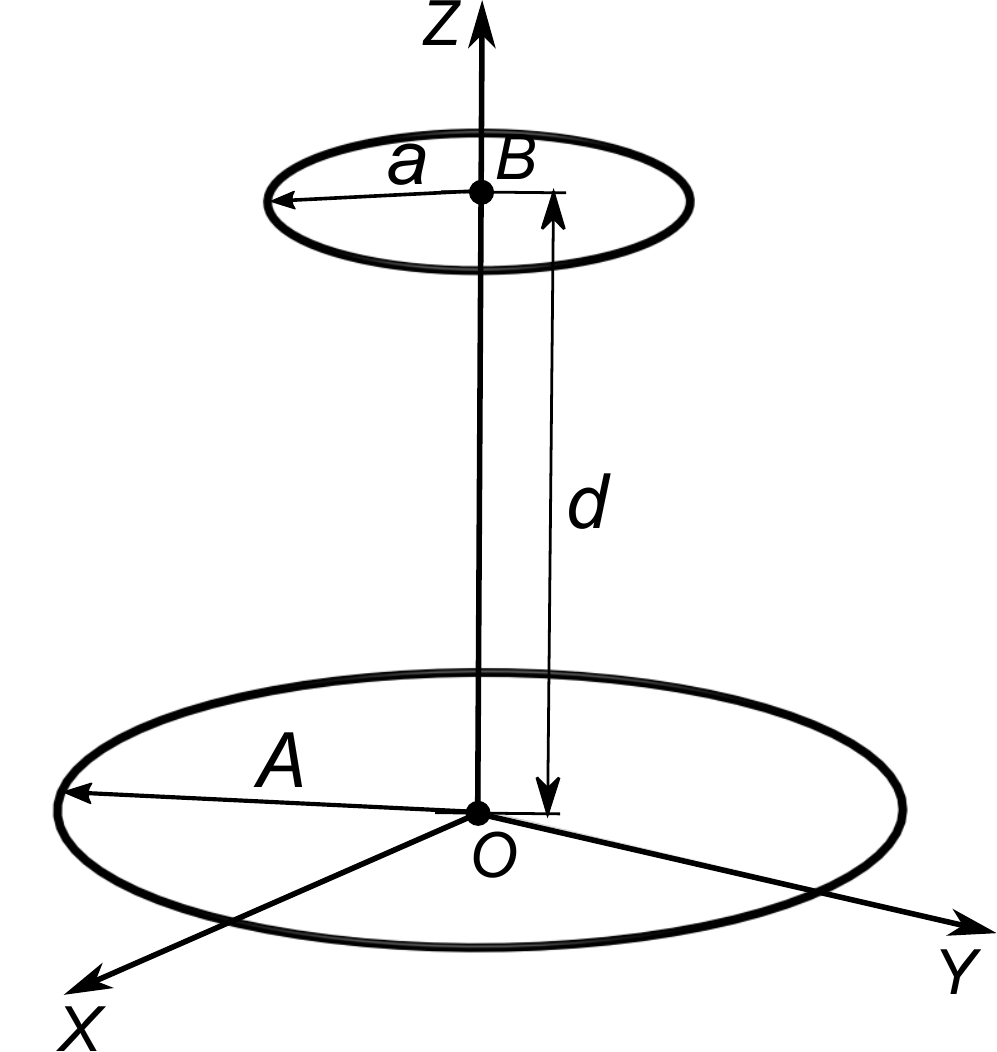}
  \caption{Geometrical scheme  of coaxial circular  filaments denoted via Grover's notation: radii $a$ and $A$ of secondary and primary coils, respectively; $d$ is the distance between the planes of circles.   }\label{fig:coaxial filaments}
\end{figure}

\subsection{ Mutual inductance of coaxial circular  filaments}
\label{sec:coaxial circular}

Let us consider the circular filaments, which are coaxial and have a distance between  their centres, as shown in Fig. \ref{fig:coaxial filaments}. Then, this case in  the notation proposed in this article is defined as $R_p=A$, $R_s=a$, the linear misalignment is $z_B=d$, $x_B=y_B=0$,  the angular misalignment (manner I, Sec. \ref{sec:Pleliminary}) is $\theta=0$ and $\eta=0$.
For the Babi\v{c} formula the linear misalignment is defined in the same way, but for angular one the parameters of the secondary circle plane must be calculated and becomes $a=0$, $b=0$ and $c=1$ (the Babi\v{c}  notation).
These parameters have the following relationship with the angle  $\theta$ and $\eta$: $a=\sin\eta\sin\theta$; $b=-\cos\eta\sin\theta$ and $c=\cos\theta$ \cite[Eq. (27), page 3597]{BabicSiroisAkyelEtAl2010}.

\subsubsection*{ Example 1 (Example 24, page 78 in Grover's book \cite{Grover2004})}
Let us suppose that two circles of radii $a$=\SI{20}{\centi\meter} and $A$=\SI{25}{\centi\meter} with their planes $d$=\SI{10}{\centi\meter} apart are given. The results of calculation are

\vspace*{1.0em}
\begin{tabular}{cccc}
  \toprule
  &Grover's book& The Babi\v{c} formula & This work, Eq. (\ref{eq:NEW FORMULA}) \\
    \midrule
   $M$, nH&   248.79 & 248.7874 & 248.7874 \\
  \toprule\label{tab:example1}
\end{tabular}

\subsubsection*{ Example 2 (Example 25, page 78 in Grover's book \cite{Grover2004})}
Two circles of radii $a$=\SI{2}{in}= \SI{5.08}{\centi\meter} and $A$=\SI{5}{in}=\SI{12.7}{\centi\meter} with their planes $d$=\SI{4}{in}=\SI{10.16}{\centi\meter} apart, the results become

\vspace*{1.0em}
\begin{tabular}{cccc}
  \toprule
  &Grover's book&  The Babi\v{c} formula& This work, Eq. (\ref{eq:NEW FORMULA}) \\
   \midrule
   $M$, nH&   18.38 & 18.3811 & 18.3811 \\
  \toprule\label{tab:example2}
\end{tabular}

\subsubsection*{ Example 3 (Example 5-4, page 215 in Kalantarov's book  \cite{Kalantarov1986})}
\label{sec:example3}
For two circles having the same radii of \SI{10.00}{\centi\meter} with their planes $d$=\SI{4}{\centi\meter} apart, the calculation shows the following

\vspace*{1.0em}
\begin{tabular}{cccc}
  \toprule
  &Kalantarov's book&  The Babi\v{c} formula& This work, Eq. (\ref{eq:NEW FORMULA})\\
   \midrule
   $M$, nH&   135.1 & 135.0739 & 135.0739 \\
  \toprule\label{tab:example3}
\end{tabular}

\subsubsection*{ Example 4 (Example 5-5, page 215 in Kalantarov's book  \cite{Kalantarov1986})}
 Circles having the same radii as in \textit{Example 3}, but their planes $d$=\SI{50}{\centi\meter} apart are given. The results are

\vspace*{1.0em}
\begin{tabular}{cccc}
  \toprule
  &Kalantarov's book&  The Babi\v{c} formula& This work, Eq. (\ref{eq:NEW FORMULA})\\
    \midrule
   $M$, nH&   1.41 & 1.4106 & 1.4106 \\
  \toprule\label{tab:example3}
\end{tabular}

\subsubsection*{ Example 5 (Example 5-6, page 224 in Kalantarov's book  \cite{Kalantarov1986})}
 Two circular coaxial filaments, radii of which are $A$=\SI{25}{\centi\meter} and $a$=\SI{20}{\centi\meter}, with their planes $d$=\SI{8}{\centi\meter} apart are given. The results are as follows

\vspace*{1.0em}
\begin{tabular}{cccc}
  \toprule
  &Kalantarov's book&  The Babi\v{c} formula & This work, Eq. (\ref{eq:NEW FORMULA})\\
   \midrule
   $M$, nH&   289.11 & 289.0404 & 289.0404 \\
  \toprule\label{tab:example3}
\end{tabular}

\begin{figure}[!t]
  \centering
  \includegraphics[width=1.8in]{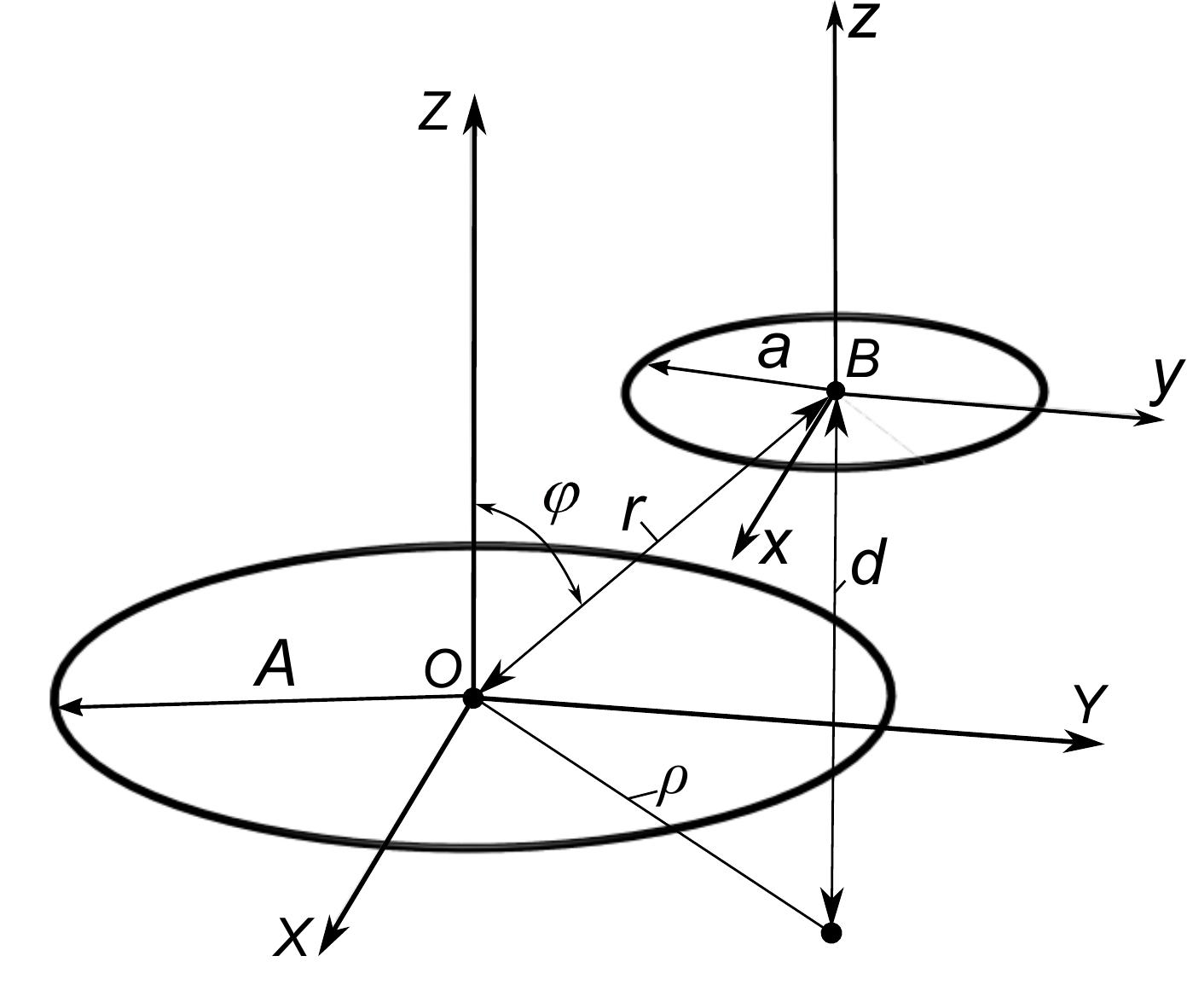}
  \caption{Geometrical scheme  of  circular  filaments with parallel axes denoted via Grover's notation: $\rho$ is the distance between axes; $r$ is the distance between the centres; $\varphi$ is the angle between the $Z$-axis and the radius vector $r$.   }\label{fig:filaments parallel axes}
\end{figure}
\subsection{ Mutual inductance of circular  filaments with parallel axes}
The scheme for calculation of the mutual inductance between circular  filaments with parallel axes is shown in Fig. \ref{fig:filaments parallel axes}. The linear misalignment in the Grover notation can be defined by $d$ is the distance between the planes of circles (the same parameter as in Sec. \ref{sec:coaxial circular}) and $\rho$ is the distance between axes or via $r$ is the distance between the centres and $\varphi$ is the angle between  the $Z$-axis and the radius vector $r$. These parameters have the following relationship to the notation defined in this article, namely, $z_B=d=r\cos\varphi$ and $\rho=\sqrt{x_B^2+y_B^2}=r\sin\varphi$. The angular misalignment is defined in the same way  as described in  Sec. \ref{sec:coaxial circular}).

\subsubsection*{ Example 6 (Example 62, page 178 in Grover's book \cite{Grover2004}))}
Two circles of radii $a=A=$\SI{15}{\centi\meter} have a distance between their centres $r=$\SI{20}{\centi\meter} and an angle $\varphi=\cos^{-1}0.8$ between the $Z$-axis and the radius  vector $r$ (please, see Fig. \ref{fig:filaments parallel axes}). Assuming that $y_B=\rho=r\sin\varphi=$\SI{12}{\centi\meter} and $z_B=r\cos\varphi=$\SI{16}{\centi\meter}, the results of calculation are as follows

\vspace*{1.0em}
\begin{tabular}{cccc}
  \toprule
  &Grover's book&  The Babi\v{c} formula & This work, Eq. (\ref{eq:NEW FORMULA})\\
   \midrule
   $M$, nH&   45.31 & 45.3342 & 45.3342 \\
  \toprule\label{tab:example6}
\end{tabular}

\subsubsection*{ Example 7 (Example 63, page 178 in Grover's book \cite{Grover2004})}
Two circles of the same diameter of $2a=2A=$\SI{48}{in}$=$\SI{121.92}{\centi\meter} are arranged so that  the distance between their planes $d=$\SI{15}{in}$=$\SI{38.1}{\centi\meter} and the distance between their axes is $\rho=$\SI{47.7}{in}$=$\SI{121.158}{\centi\meter} (please, see Fig. \ref{fig:filaments parallel axes}). Thus, we have $R_p=R_s=$\SI{60.96}{\centi\meter}, $y_B=\rho$ and $z_B=d$, the results of calculation are as follows

\vspace*{1.0em}
\begin{tabular}{cccc}
  \toprule
  &Grover's book& The Babi\v{c} formula& This work, Eq. (\ref{eq:NEW FORMULA})\\
   \midrule
   $M$, nH&   $-24.56$ & $-24.5728$ & $-24.5728$ \\
  \toprule\label{tab:example7}
\end{tabular}

\subsubsection*{ Example 8 (Example 65, page 183 in Grover's book \cite{Grover2004})}
Two circles with radii of $A=$\SI{10}{\centi\meter} and $a=$\SI{8}{\centi\meter} have the distance between their centres $r=$\SI{50}{\centi\meter} and an angle of $\cos\varphi=0.4$ (please, see Fig. \ref{fig:filaments parallel axes}). Hence, we have $R_p=A$ and $R_s=a$, assuming that $y_B=r\sin\varphi=$\SI{45.83}{\centi\meter} and $z_B=r\cos\varphi=$\SI{20.0}{\centi\meter}, the results of calculation are as follows

\vspace*{1.0em}
\begin{tabular}{cccc}
  \toprule
  &Grover's book&  The Babi\v{c} formula & This work, Eq. (\ref{eq:NEW FORMULA})\\
   \midrule
   $M$, nH&   $-0.2480$ & $-0.24828$ & $-0.24828$ \\
  \toprule\label{tab:example8}
\end{tabular}
\subsubsection*{ Example 9 (Example 66, page 184 in Grover's book \cite{Grover2004})}
Two circles with radii of $A=$\SI{10}{\centi\meter} and $a=$\SI{8}{\centi\meter} have the distance between their centres $r=$\SI{20}{\centi\meter} and an angle of $\cos\varphi=0.6$ (please, see Fig. \ref{fig:filaments parallel axes}), to find the mutual inductance between these circles. Hence, we have $R_p=A$ and $R_s=a$, assuming that $y_B=r\sin\varphi=$\SI{16.0}{\centi\meter} and $z_B=r\cos\varphi=$\SI{12.0}{\centi\meter}, the results of calculation are as follows

\vspace*{1.0em}
\begin{tabular}{cccc}
  \toprule
  &Grover's book&  The Babi\v{c} formula& This work, Eq. (\ref{eq:NEW FORMULA}) \\
   \midrule
   $M$, nH&   $4.405$ & $4.465$ & $4.465$ \\
  \toprule\label{tab:example9}
\end{tabular}

\subsubsection*{ Example 10 (Example 5-8, page 231 in Kalantarov's book  \cite{Kalantarov1986})}
 Two circular  filaments of the same radius of $A=a=$\SI{5}{\centi\meter} are arranged that the distance between their centres is  $r$=\SI{40}{\centi\meter} and  an angle of $\cos\varphi=0.4$. Hence, we have $R_p=R_s=$\SI{5}{\centi\meter}, assuming that $y_B=r\sin\varphi=$\SI{36.66}{\centi\meter} and $z_B=r\cos\varphi=$\SI{16.0}{\centi\meter}, the results are as follows

\vspace*{1.0em}
\begin{tabular}{cccc}
  \toprule
  &Kalantarov's book&  The Babi\v{c} formula  & This work, Eq. (\ref{eq:NEW FORMULA})\\
   \midrule
   $M$, nH&   $-0.049$  & $-0.048963$  & $-0.048963$ \\
  \toprule\label{tab:example10}
\end{tabular}

\subsubsection*{ Example 11 (Example 5-9, page 233 in Kalantarov's book  \cite{Kalantarov1986})}
 Two circular  filaments of radii of $A=$\SI{10}{\centi\meter} and $a=$\SI{5}{\centi\meter} are arranged that the distance between their centres is  $r=$\SI{20}{\centi\meter} and  an angle of $\cos\varphi=0.8$. Hence, we have $R_p=$\SI{10}{\centi\meter} and $R_s=$\SI{5}{\centi\meter}, assuming that $y_B=r\sin\varphi=$\SI{36.66}{\centi\meter} and $z_B=r\cos\varphi=$\SI{16.0}{\centi\meter}, the results are as follows

\vspace*{1.0em}
\begin{tabular}{cccc}
  \toprule
  &Kalantarov's book&  The Babi\v{c} formula& This work, Eq. (\ref{eq:NEW FORMULA})\\
   \midrule
   $M$, nH&   $2.95$  & $3.0672$  & $3.0672$ \\
  \toprule\label{tab:example11}
\end{tabular}

\subsubsection*{ Example 12 (Example 5-10, page 234 in Kalantarov's book  \cite{Kalantarov1986})}
 Two circular  filaments of radii of $A=$\SI{20}{\centi\meter} and $a=$\SI{4}{\centi\meter} are arranged that the distance between their centres is  $r=$\SI{2}{\centi\meter} and  an angle of $\cos\varphi=0.66$. Hence, we have $R_p=$\SI{20}{\centi\meter} and $R_s=$\SI{4}{\centi\meter}, assuming that $y_B=r\sin\varphi=$\SI{1.5}{\centi\meter} and $z_B=r\cos\varphi=$\SI{1.32}{\centi\meter}, the results become as follows

\vspace*{1.0em}
\begin{tabular}{cccc}
  \toprule
  &Kalantarov's book&  The Babi\v{c} formula& This work,Eq. (\ref{eq:NEW FORMULA}) \\
   \midrule
   $M$, nH&   $15.99$  & $15.9936$  & $15.9936$ \\
  \toprule\label{tab:example12}
\end{tabular}

\begin{figure}[!t]
  \centering
  \includegraphics[width=1.8in]{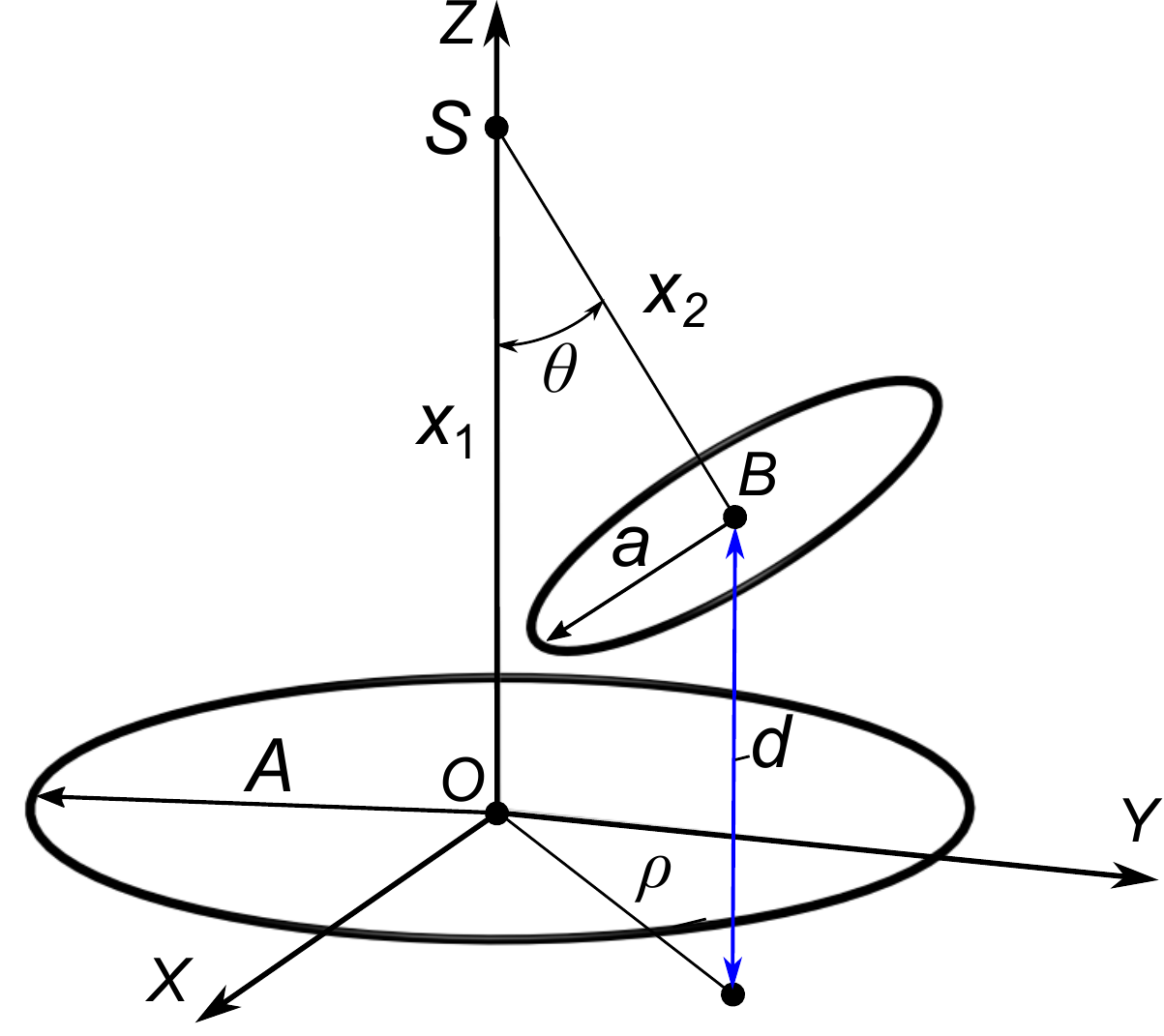}
  \caption{Geometrical scheme  of inclined circular  filaments with intersect axes denoted via Grover's notation: $x_1$ and $x_2$  are the distances from $S$,  $\theta$ is the angle of inclination of the axes.   }\label{fig:filaments intersect axes}
\end{figure}
\subsection{ Mutual inductance of inclined circular  filaments with intersect axes}

The general scheme of the arrangement of two inclined circular filaments whose axes intersect for calculation of mutual inductance is shown in Fig. \ref{fig:filaments intersect axes}. In Grover's notation, we have $A$ and $a$ are radii of circular filaments and $S$ is the point of intersection of the circles axes.  The centres of circles are at the distances $x_1$ and $x_2$ from $S$ of the primary and secondary circle, respectively.  $\theta$ is the angle of inclination of the axes. Also, the following relationships are true, namely, $d=x_1-x_2\cos\theta$ and $\rho=x_2\sin\theta$. Thus, the linear misalignment in the notation of this paper is defined again through  $z_B=d$, $\rho=\sqrt{x_B^2+y_B^2}$ and the angular misalignment is defined by the angle $\theta$, but $\eta$ is equal to zero. For the Babi\v{c} formula, the angular misalignment is  defined by the  parameters of the secondary circle plane as follows: $a=0$; $b=-\sin\theta$ and $c=\cos\theta$.

From the general scheme shown in Fig. \ref{fig:filaments intersect axes}, two particular cases can be recognized. Namely, the first case is corresponded to  concentric circles, when  $x_1=x_2=0$ and the second case is corresponded to circular filaments whose axes intersect at the centre of one of the circle, when $x_2=0$. For the first case, to calculate the mutual inductance, the angle  $\theta$ and radii of circles must be known. For the second particular case, in addition to the distance $d$ between the centre $B$ of the secondary circle and the plane of the primary circle must be given.

\subsubsection*{ Example 13 (Example 5-7, page 227 in Kalantarov's book  \cite{Kalantarov1986})}
 Two circular  filaments of radii of $A=$\SI{10}{\centi\meter} and $a=$\SI{2.5}{\centi\meter} are concentric and   an angle of inclination of the plane of the secondary circle is $\theta=$\SI{60}{\degree}. Hence, assuming that $x_B=y_B=z_B=0$, the calculation of mutual inductance shows

\vspace*{1.0em}
\begin{tabular}{cccc}
  \toprule
  &Kalantarov's book & The Babi\v{c} formula& This work, Eq. (\ref{eq:NEW FORMULA})\\
   \midrule
   $M$, nH&   $6.044$  & $6.0431$  & $6.0431$ \\
  \toprule\label{tab:example13}
\end{tabular}

\subsubsection*{ Example 14 (Example 69, page 194 in Grover's book \cite{Grover2004})}
Two concentric circles with radii of $A=$\SI{20}{\centi\meter} and $a=$\SI{14}{\centi\meter} are arranged that  an angle of inclination of the plane of the secondary circle is $\cos\theta=0.3$. Hence, assuming that $x_B=y_B=z_B=0$ and $\theta=$\SI{72.5424}{\degree}, the results of calculation are as follows

\vspace*{1.0em}
\begin{tabular}{cccc}
  \toprule
  &Grover's book&  The Babi\v{c} formula & This work, Eq. (\ref{eq:NEW FORMULA})\\
   \midrule
   $M$, nH&   $47.44$ & $47.4431$ & $47.4431$ \\
  \toprule\label{tab:example14}
\end{tabular}

\subsubsection*{ Example 15 (Example 70, page 194 in Grover's book \cite{Grover2004})}
Two circles with radii of $A=$\SI{10}{in}=\SI{25.4}{\centi\meter} and $a=$\SI{3}{in}=\SI{7.62}{\centi\meter} are arranged that  an angle of inclination of the plane of the secondary circle is $\cos\theta=0.4$ and a distance $d=$\SI{3}{in}=\SI{7.62}{\centi\meter}. Hence, assuming that $x_B=y_B=0$, $z_B=d$ and $\theta=$\SI{66.4218}{\degree}, the results of calculation are as follows

\vspace*{1.0em}
\begin{tabular}{cccc}
  \toprule
  &Grover's book&  The Babi\v{c} formula & This work, Eq. (\ref{eq:NEW FORMULA})\\
   \midrule
   $M$, nH&   $15.543$ & $15.5435$ & $15.5435$ \\
  \toprule\label{tab:example15}
\end{tabular}

\subsubsection*{ Example 16 (Example 71, page 201 in Grover's book \cite{Grover2004})}
Two circles of radii $A=$\SI{20.0}{\centi\meter} and $a=$\SI{10.0}{\centi\meter} are considered with the centre of one on the axis of the other and a distance, $d$, of \SI{20.0}{\centi\meter} between the centres. The axes are to be inclined at an angle, $\theta$, of \SI{30}{\degree}.
The results of calculation show

\vspace*{1.0em}
\begin{tabular}{cccc}
  \toprule
  &Grover's book&  The Babi\v{c} formula & This work, Eq. (\ref{eq:NEW FORMULA})\\
   \midrule
   $M$, nH&   $29.436$ & $29.4365$ & $29.4365$ \\
  \toprule\label{tab:example16}
\end{tabular}

\subsubsection*{ Example 17 (Example 5-11, page 235 in Kalantarov's book  \cite{Kalantarov1986})}
 Two circular  filaments have radii of $A=$\SI{10}{\centi\meter} and $a=$\SI{8}{\centi\meter}. The axis the primary circle is crossed through the centre of the secondary circle at a distance $d$ of \SI{8}{\centi\meter} between their centres. The axes are to be inclined at an angle, $\cos\theta=0.7$. Hence, assuming that $x_B=y_B=0$, $z_B=d$ and an angle of \SI{45.5730}{\degree}, the results of calculation are

\vspace*{1.0em}
\begin{tabular}{cccc}
  \toprule
  &Kalantarov's book&  The Babi\v{c} formula  & This work, Eq. (\ref{eq:NEW FORMULA})\\
   \midrule
   $M$, nH&   $23.2$  & $24.3794$  & $24.3794$ \\
  \toprule\label{tab:example17}
\end{tabular}

\subsubsection*{ Example 18 (Example 73, page 204 in Grover's book \cite{Grover2004})}
Two circles of radii $A=$\SI{16.0}{\centi\meter} and $a=$\SI{10.0}{\centi\meter} are considered to be intersect the axes at point  $S$ in such a way that distances $x_1$ and $x_2$ are to be  \SI{20.0}{\centi\meter} and  \SI{5.0}{\centi\meter}, respectively. An angle of inclination between axes is $\cos\theta=0.5$.   Hence, assuming that $x_B=0$, $y_B=$\SI{4.3301}{\centi\meter}, $z_B=$\SI{17.5}{\centi\meter} and an angle of \SI{60.0}{\degree}, we have

\vspace*{1.0em}
\begin{tabular}{cccc}
  \toprule
  &Grover's book& The Babi\v{c} formula & This work, Eq. (\ref{eq:NEW FORMULA})\\
   \midrule
   $M$, nH&   $13.612$ & $13.6113$ & $13.6113$ \\
  \toprule\label{tab:example18}
\end{tabular}
\begin{table}[!b]
  \caption{Calculation of mutual inductance for Example 19 }\label{tab:Example19}
  \centering
\begin{tabular}{lccc}
 \toprule
  $\eta$ & The Grover formula, &  The Babi\v{c} & This work,\\
   &  \cite[Eq. (179)]{Grover2004} &formula,   \cite[Eq. (24)]{BabicSiroisAkyelEtAl2010}&Eq. (\ref{eq:NEW FORMULA})\\
   & $M$, nH& $M$, nH &$M$, nH\\
  \midrule
  0 & 13.6113&13.6113& 13.6113\\
  $\pi/6$ & 14.4688&14.4688& 14.4688\\
   $\pi/4$ & 15.4877&15.4877& 15.4877\\
   $\pi/4$ & 16.8189&16.8189& 16.819\\
    $\pi/2$ & 20.0534&20.0534& 20.0534\\
     $2\pi/3$ & 23.3252&23.3252& 23.3252\\
      $3\pi/4$ & 24.6936&24.6936&24.6936\\
       $5\pi/6$ & 25.7493&25.7493&25.7493\\
        $\pi$ & 26.6433&26.6433& 26.6433\\
        $7\pi/6$ & 25.7493&25.7493& 25.7493\\
        $5\pi/4$ & 24.6936& 24.6936& 24.6936\\
        $4\pi/3$ & 23.3253&23.3253& 23.3252\\
        $3\pi/2$ & 20.0534&20.0534& 20.0534\\
        $5\pi/3$ & 16.8189&16.8189& 16.819\\
        $7\pi/4$ & 15.4877&15.4877& 15.4877\\
        $11\pi/6$ & 14.4688 &14.4688 & 14.4688 \\
        $2\pi$ & 13.6113&13.6113& 13.6113\\
 \toprule
\end{tabular}
\end{table}

\subsection{ Mutual inductance of circular  filaments arbitrarily positioned in the space}
The validation of the developed formulas (\ref{eq:NEW FORMULA}) and (\ref{eq:Singular case}) for the general case, when the angular misalignment is defined through the angle $\theta$ and $\eta$ as shown in Fig. \ref{fig:angular position}(a) in an range from 0 to 360\si{\degree}, the examples from the Babi\v{c} article  \cite{BabicSiroisAkyelEtAl2010} were used. Also, we utilized the Matlab functions with the Grover formula \cite[page 207, Eq. (179)]{Grover2004} and the Babi\v{c} formula \cite[page 3593, Eq. (24)]{BabicSiroisAkyelEtAl2010} implemented by F. Sirois and S. Babi\v{c}.

\subsubsection*{ Example 19 (Example 12, page 3597 in the Babi\v{c} article \cite{BabicSiroisAkyelEtAl2010})}
Using the geometrical arrangement as in Example 18 (two circles with radii $A=$\SI{16.0}{\centi\meter} and $a=$\SI{10.0}{\centi\meter} and the centre of the secondary circle is located at $x_B=0$, $y_B=$\SI{4.3301}{\centi\meter}, $z_B=$\SI{17.5}{\centi\meter} and the angle, $\theta$ of \SI{60.0}{\degree}), but the angle $\eta$ is varied in a range from 0 to 360\si{\degree}. The results of calculation are summed up in Table \ref{tab:Example19}. Analysis of  Table \ref{tab:Example19} shows that the developed formula (\ref{eq:NEW FORMULA}) works identically to the Grover and Babi\v{c} formula.

\subsubsection*{ Example 20 (Example 11, page 3596 in the Babi\v{c} article \cite{BabicSiroisAkyelEtAl2010})}

Let us consider two circular filaments having radii of $R_p=$\SI{40}{\centi\meter} and $R_s=$\SI{10}{\centi\meter}, which are  mutually perpendicular to each
other that angles of $\eta=$\SI{0}{} and $\theta=$\SI{90.0}{\degree}. The centre of the secondary circle has the following coordinates:  $x_B=0$, $y_B=$\SI{20}{\centi\meter}, and $z_B=$\SI{10}{\centi\meter}.  The problem illustrates the application of new formula (\ref{eq:Singular case}). The results are

\vspace*{1.0em}
\begin{tabular}{cccc}
  \toprule
  &The Grover formula&  The Babi\v{c} formula& This work,Eq. (\ref{eq:Singular case}) \\
   \midrule
   $M$, nH&   $-10.73$ & $-10.73$ & $-10.7272$ \\
  \toprule\label{tab:example20}
\end{tabular}

\begin{figure}[!t]
  \centering
  \includegraphics[width=3.0in]{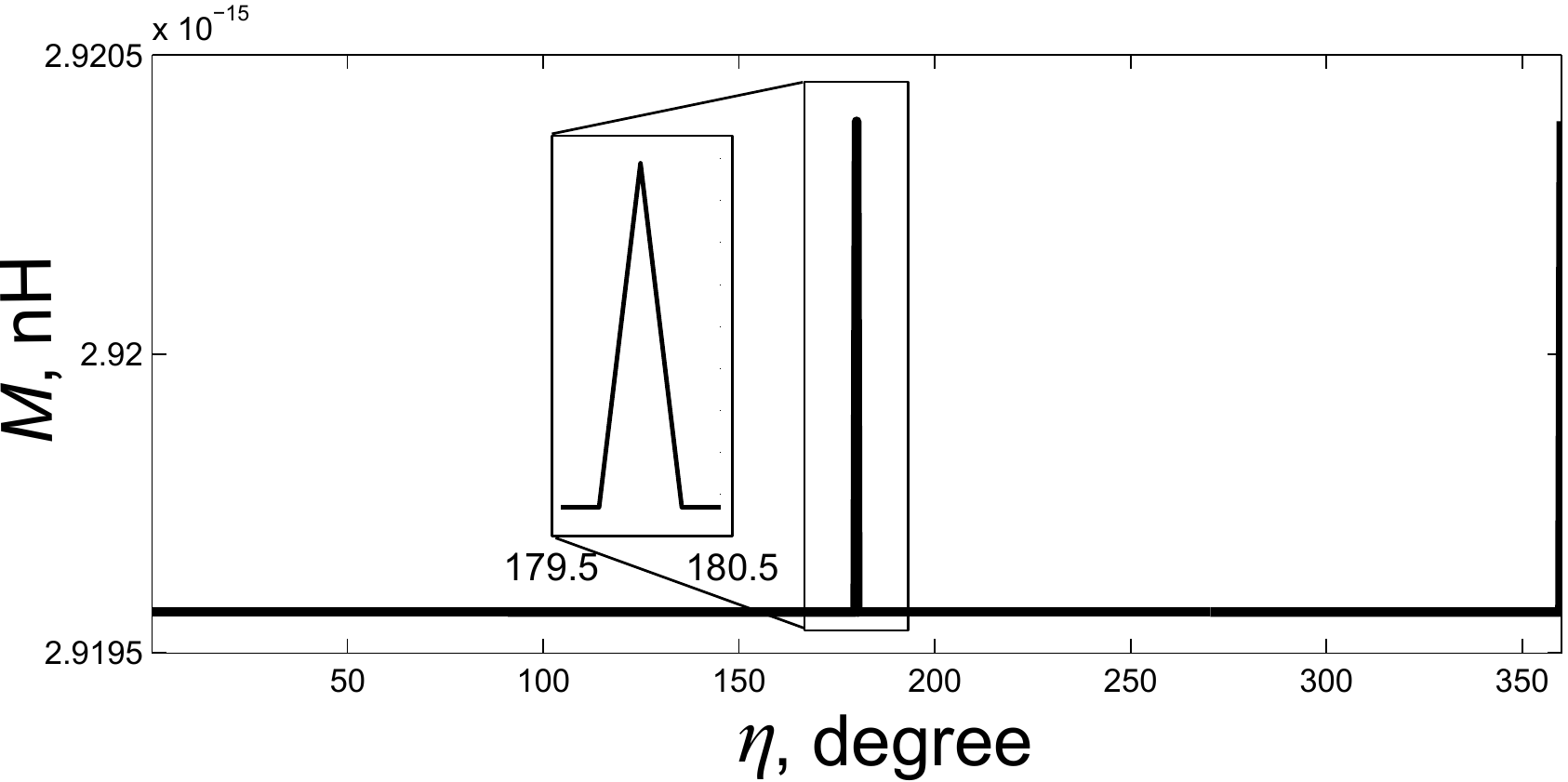}
  \caption{Distribution of the error of the Babi\v{c} formula in dependent on changing the $\eta$-angle within a range $0<\eta\leq360^{o}$ for Example 21 ($x_B=y_B=z_B=0$).   }\label{fig:Babic error in origin}
\end{figure}
\subsubsection*{ Example 21 }
Now we again apply formula (\ref{eq:Singular case}) to the problem considered in Example 20, but in this case the centre of the secondary coil is located at origin, thus  $x_B=y_B=z_B=0$. Hence, we have

\vspace*{1.0em}
\begin{tabular}{cccc}
  \toprule
  &The Grover formula&  The Babi\v{c} formula& This work, Eq. (\ref{eq:Singular case})\\
   \midrule
   $M$, nH&   $NaN$\footnotemark[1] & $NaN$ & $0$ \\
  \toprule
  \label{tab:example21}
\end{tabular}
\footnotetext[1]{Not-a-Number}

Thus, the calculation shows that the Babi\v{c} and Grover formula gives an indeterminate results, but developed formula (\ref{eq:Singular case}) equals explicitly zero as expected for this case.
Then, rotating the angle $\eta$ in a range $0<\eta\leq360^{o}$, we reveal  that the calculation of mutual inductance performed by  developed formula (\ref{eq:Singular case}) shows zero within this range of the $\eta$-angle, but the Babi\v{c} formula demonstrates a small error, which is not exceeded  $M=$\SI{2.9205e-15}{\nano\henry} and distributed with the $\eta$-angle as shown in Figure \ref{fig:Babic error in origin}.

\begin{figure}[!t]
  \centering
  \includegraphics[width=2.8in]{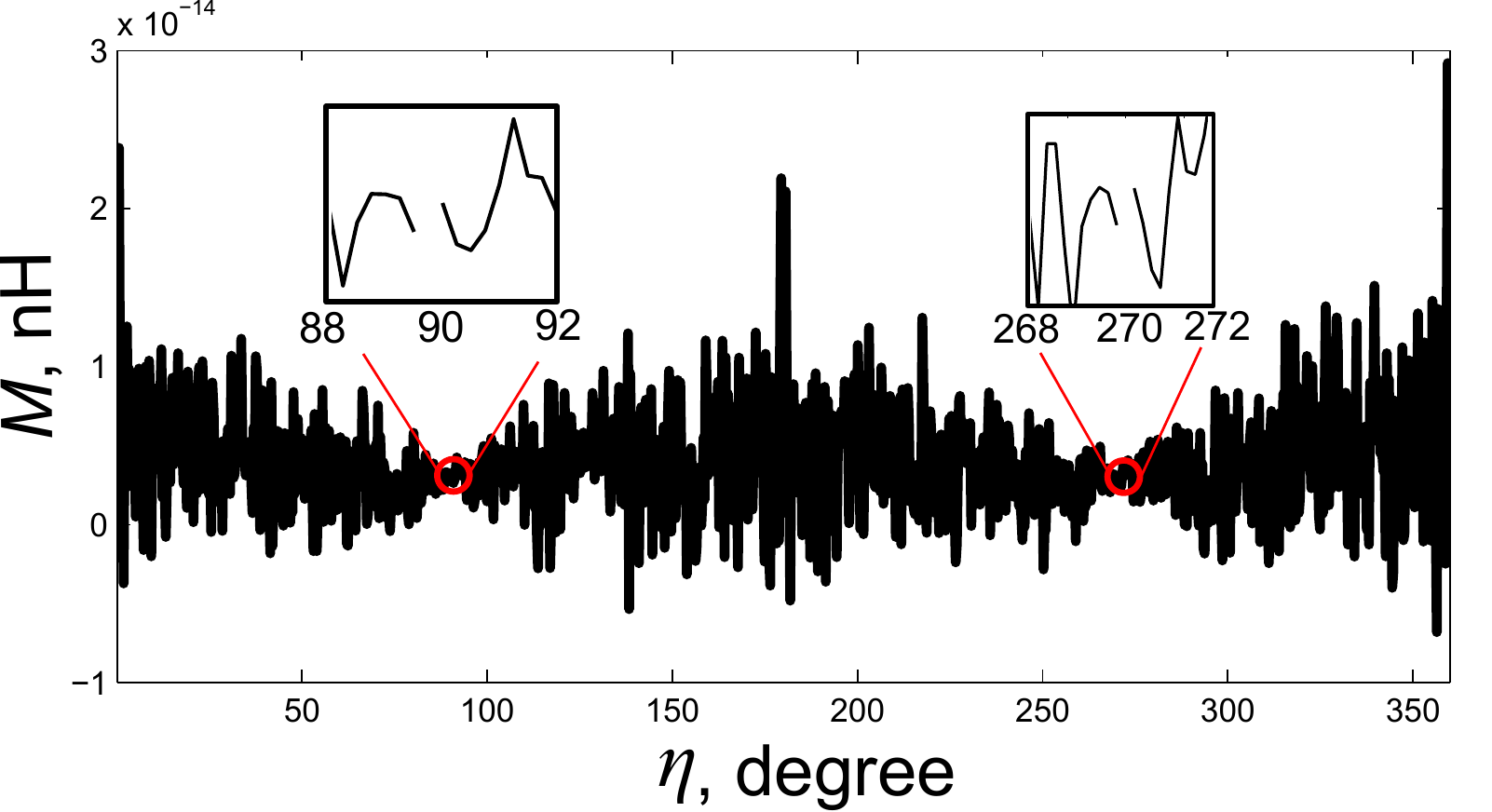}
  \caption{Chaotic distribution of the error of the Babi\v{c} formula in dependent on changing the $\eta$-angle within a range $0<\eta\leq360^{o}$ for Example 22 ($x_B=y_B=$\SI{10}{\centi\meter} and  $z_B=0$): the scaled-up images show the interruption of continuity of the curve at $\eta=$\SI{90}{\degree} and \SI{270}{\degree}.   }\label{fig:Babic error in x y}
\end{figure}
\subsubsection*{ Example 22 }
Let us consider  mutually perpendicular circles (angles of $\theta=$\SI{90.0}{\degree} and $\eta=$\SI{0}{}) having the same radii as in Example 20, but the centre of the secondary coil occupies a position on the $XOY$-surface with the following coordinates   $x_B=y_B=$\SI{10}{\centi\meter} and $z_B=0$. Results of calculation are

\vspace*{1.0em}
\begin{tabular}{cccc}
  \toprule
  &The Grover formula&  The Babi\v{c} formula & This work, Eq. (\ref{eq:Singular case})\\
   \midrule
   $M$, nH&   $4.013\times10^{-15}$& $2.416\times10^{-15}$ & $0$ \\
  \toprule
  \label{tab:example21}
\end{tabular}

\begin{figure}[!t]
  \centering
  \includegraphics[width=2.8in]{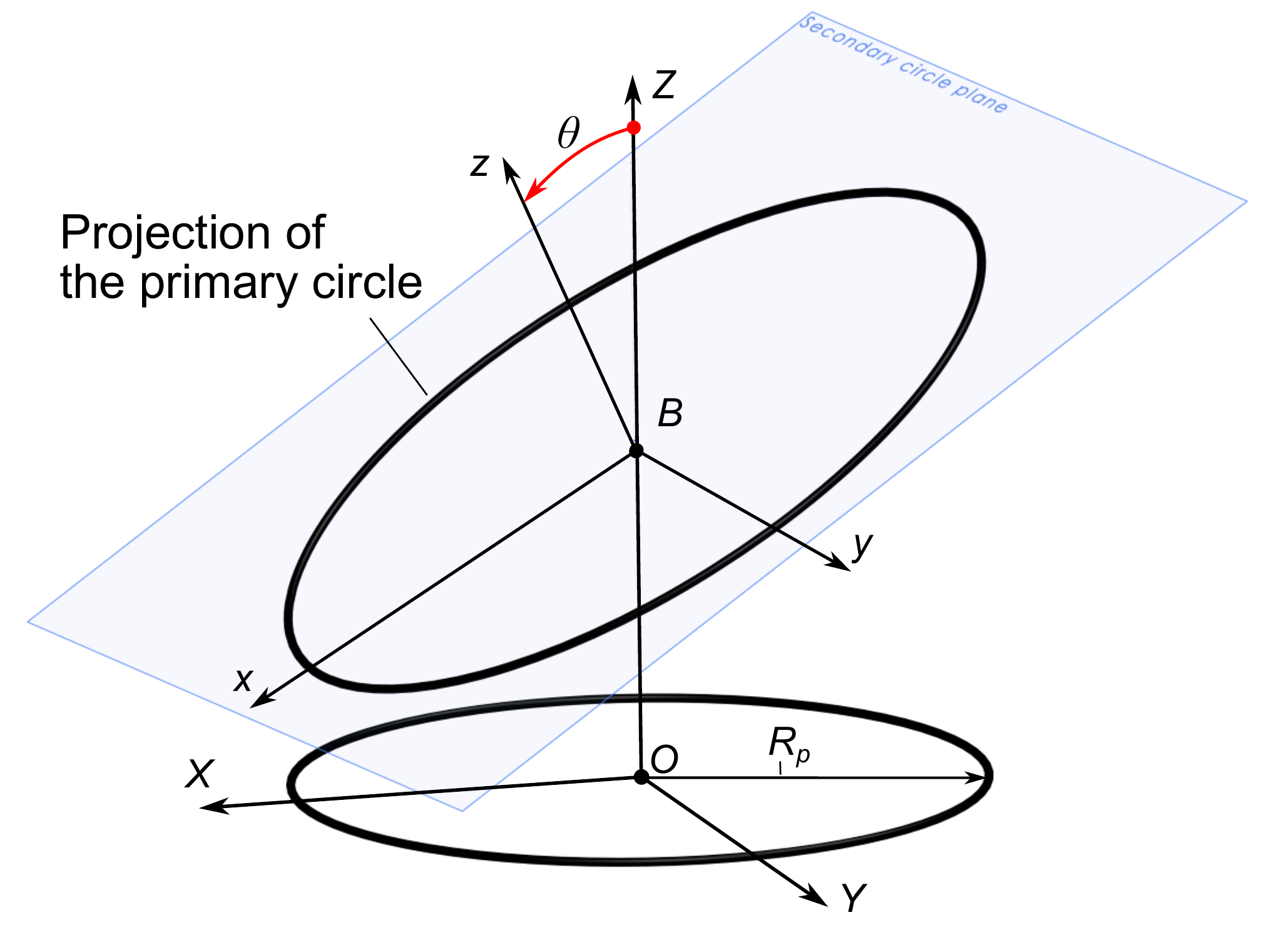}
  \caption{Geometrical scheme for calculation of mutual inductance  between a primary circular filament and its projection.  The angular misalignment is given by an angle of $\theta$, while the linear misalignment by the coordinate $z_B$.}   \label{fig:projection}
\end{figure}
Thus, the calculation expresses that the Babi\v{c} and Grover formula gives the small errors, but developed formula (\ref{eq:Singular case}) shows, explicitly,  zero.
Then, again let us rotate the angle $\eta$ in a range $0<\eta\leq360^{o}$,  the calculation of mutual inductance performed by  developed formula (\ref{eq:Singular case}) reveals zero within this range of the $\eta$-angle, but the Babi\v{c} formula demonstrates the chaotic distribution of the small error of calculation as shown in Fig. \ref{fig:Babic error in x y}, which is in a range from  \SI{-0.7e-14}{} to  \SI{2.921e-14}{\nano\henry} and at the $\eta$-angle of \SI{90}{\degree} and \SI{270}{\degree} the calculation of mutual inductance is indeterminate (see, the scaled-up images of Fig. \ref{fig:Babic error in x y}, which present the interruption of continuity  at $\eta=$\SI{90}{\degree} and \SI{270}{\degree}).

\subsection{ Mutual inductance  between a primary circular filament and its projection on a tilted plane}
In this section new formula (\ref{eq:PROJECTOIN FORMULA}) for calculation of mutual inductance between a primary circular filament and its projection on a tilted plane is validated by comparison with the calculation performed via    the \textit{FastHenry} software \cite{KamonTsukWhite1994}. The angular misalignment is given by the angle $\theta$ and $\eta$, while the linear misalignment is defined by the coordinate $z_B$ of the point $B$ crossing the tilted plane and the $Z$-axis. Fig. \ref{fig:projection} shows  a geometrical scheme for the calculation. The shown arrangement of the primary circle and its projection on the tilted plane corresponds to  a particular case, when  $\eta=0$.  Worth noting that  the $\eta$-angle has no effect on the result of  the calculation of the mutual inductance.

\begin{table}[!t]
  \caption{Calculation of mutual inductance for Example 23 }\label{tab:Example23}
  \centering
\begin{tabular}{lcc}
 \toprule
  $\theta$ &    This work, Eq. (\ref{eq:PROJECTOIN FORMULA})& \textit{FastHenry} \cite{KamonTsukWhite1994}\\
  &$M$, nH&$M$, nH\\
  \midrule
  0 & 135.0739& 135.076\\
   \SI{10}{\degree} & 142.0736& 142.086\\
   \SI{15}{\degree} & 153.3233 & 153.298\\
 \toprule
\end{tabular}
\end{table}
\subsubsection*{ Example 23 }
\label{sec:example23}
Let us consider primary circle having a radius of \SI{10.00}{\centi\meter} and a tilting plane  crosses the $Z$-axis at the point $z_B$=\SI{4}{\centi\meter}. When a tilting angle of zero, then the geometry and arrangement corresponds to Example 3 (Example 5-4, page 215 in Kalantarov's book) for the case of two coaxial circles with the same radii. We calculate the mutual inductance for three values of a tilted angle at \SI{0}, \SI{10}{\degree} and  \SI{15}{\degree}. The results of calculation are shown in Table \ref{tab:Example23}.

Although,  there is the small deviation  between results obtained with the \textit{FastHenry} software  and analytical formula (\ref{eq:PROJECTOIN FORMULA}), but this deviation is not significant and can be explained by the fact that the circles in  the \textit{FastHenry} software are divided into straight segments with a finite cross section in comparing with the analytical formula where the  circles have no segments and a cross section.  We are concluding the validity of developed formula (\ref{eq:PROJECTOIN FORMULA}).

\begin{table}[!b]
  \caption{Calculation of mutual inductance for Example 24 }\label{tab:Example24}
  \centering
\begin{tabular}{lcc}
 \toprule
  $\theta$ &    This work, Eq. (\ref{eq:PROJECTOIN FORMULA})& \textit{FastHenry} \cite{KamonTsukWhite1994}\\
  &$M$, nH&$M$, nH\\
  \midrule
  0 & 1.4106& 1.3761\\
  \SI{5}{\degree} & 1.4117& 1.3844\\
   \SI{10}{\degree} & 1.4151& 1.3933\\
   \SI{15}{\degree} & 1.421 & 1.4031\\
     \SI{20}{\degree} & 1.4298 & 1.4120\\
    \SI{25}{\degree} & 1.4422 & 1.4202\\
    \SI{30}{\degree} & 1.4594 & 1.4268\\
      \SI{35}{\degree} & 1.4831 & 1.4336\\
       \SI{40}{\degree} & 1.5161 & 1.4574\\
        \SI{45}{\degree} & 1.5631 & 1.5230\\
        \SI{50}{\degree} & 1.6329 & 1.6180\\
        \SI{55}{\degree} & 1.7425& 1.7273\\
        \SI{60}{\degree} & 1.9299& 1.8765\\
          \SI{65}{\degree} & 2.2971& 2.2416\\
          \SI{70}{\degree} & 3.2127& 3.1806\\
          \SI{75}{\degree} & 7.1274& 7.1679\\
          \toprule
\end{tabular}
\end{table}

\subsubsection*{ Example 24 }
\label{sec:example24}
In this last example, we increase a  distance  between the centre of the primary circle with a radius of \SI{10.00}{\centi\meter} and a tilting plane to  $z_B$=\SI{50}{\centi\meter}. Hence, a range of the tilted angle becomes larger then in example 23. Note that for zero tilting angle  the geometry of the considered problem corresponds to example 4 (Example 5-5, page 215 in Kalantarov's book). The results of calculation are shown in Table \ref{tab:Example24}. Analysis of Table \ref{tab:Example24} shows a good agreement between the calculations, which confirms the validity of developed formula (\ref{eq:PROJECTOIN FORMULA}).

\section{Conclusion}

We derived and validated new formulas (\ref{eq:NEW FORMULA}) and (\ref{eq:Singular case}) for calculation of the mutual inductance between two  circular filaments arbitrarily oriented with respect to each other.
These analytic formulas have been developed based on  the Kalantarov-Zeitlin method, which showed that the calculation of mutual inductance between a circular primary filament and any other secondary filament having an arbitrary shape and any desired position with respect to the primary filament is reduced to a line integral. In particular, the developed formula  (\ref{eq:Singular case}) provides a solution for the singularity issue arising in Grover's and Babi\v{c}' formulas for the case when the planes of the primary and secondary circular filaments are mutually perpendicular.

Moreover, a curious reader  can already recognize that    formula (\ref{eq:Singular case}) can be applied for calculation of the mutual inductance between the circular filament and a line, position of which with respect to the circle is defined through the linear and angular misalignment. For this reason, in Eq. (\ref{eq: Z}) we assume that   $\bar{z}_{\lambda}=\bar{z}_B$ and  formula (\ref{eq:Singular case}) is integrated only from $-1$ to $1$.  This fact proves again  the efficiency and flexibility of the
Kalantarov-Zeitlin method.

 The advantages of the Kalantarov-Zeitlin method allow us to extend immediately the application of the obtained result to a case of the calculation of the mutual inductance between a primary circular filament and its projection on a tilted plane and  to furnish this case via formula (\ref{eq:PROJECTOIN FORMULA}). For instance, this particular case appears in micro-machined inductive suspensions and has a direct practical application in studying their stability  and pull-in dynamics.

 New
developed formulas have been successfully validated through a number of
examples available in the literature. Also,  the direct comparison the results of calculation with
the numerical results obtained by utilizing the \textit{FastHenry} software  shows a good agreement.
Besides, the obtained formulas can be easily programmed, they are
intuitively understandable for application.

\section*{Acknowledgment}
Kirill Poletkin acknowledges with thanks to Prof. Ulrike Wallrabe for the continued support of his research. Also, Kirill Poletkin acknowledges with thanks the support from German Research Foundation (Grant KO 1883/26-1).

%
\appendix

\begin{figure}[!b]
  \centering
  \includegraphics[width=2.5in]{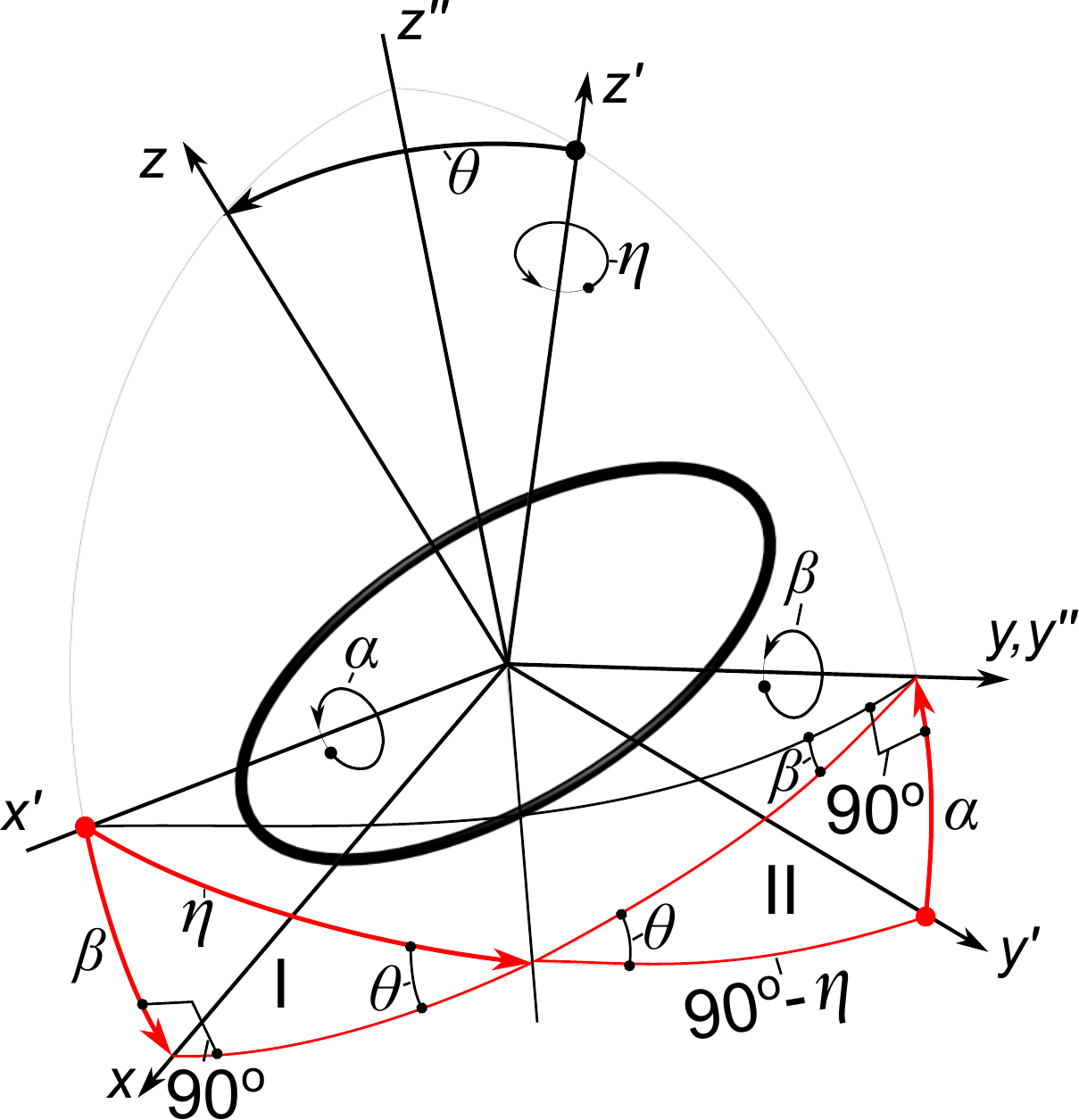}
  \caption{The relationship between the angles of  two manners for determining angular misalignment of the secondary circle: I and II are denoted for two spherical triangles highlighted by arcs in red color.   }\label{fig:angular misalignment}
\end{figure}
\section{Determination of angular position of the secondary circular filament}
\label{app:determination}
The angular position of the secondary circle can be defined through the pair of angle $\theta$ and $\eta$ corresponding to manner I and the angle   $\alpha$ and $\beta$ manner II.  The relationship between two pairs of angles can be determined via two spherical triangles denoted in Roman number I and II as shown in Fig. \ref{fig:angular misalignment}. According to the law of sines, for spherical triangle I we can write the following relationship:
\begin{equation}\label{eq: triangleI}
\frac{\sin\eta}{\sin\pi/2}=\frac{\sin\beta}{\sin\theta}.
\end{equation}
 For spherical triangle II, we have
\begin{equation}\label{eq: triangleII}
\frac{\sin(\pi/2-\eta)}{\sin(\pi/2-\beta)}=\frac{\sin\alpha}{\sin\theta}.
\end{equation}
Accounting for (\ref{eq: triangleI}) and (\ref{eq: triangleII}), the final set determining the relationship between two pairs of angles becomes as
 \begin{equation}\label{eq:angles pair}
  \left\{\begin{array}{l}
   \sin\beta=\sin\eta\sin\theta;\\
   \cos\beta\sin\alpha=\cos\eta\sin\theta.
  \end{array}\right.
\end{equation}
%

\section{Presentation of developed formulas via the pair of angles $\alpha$ and $\beta$}
\label{app:formulas}
Using set (\ref{eq:angles pair}), we can write the following equations:
 \begin{equation}\label{eq:new pair}
  \left\{\begin{array}{l}
   \cos^2\theta=\cos^2\beta(1+\sin^2\alpha);\\
     \sin^2\theta=\sin^2\beta+\cos^2\beta\sin^2\alpha;\\
  {\displaystyle   \tan^2\theta=\frac{\sin^2\beta+\cos^2\beta\sin^2\alpha}{\cos^2\beta(1+\sin^2\alpha)};}\\
    {\displaystyle \cos^2\eta=\cos^2\beta\sin^2\alpha/{\sin^2\theta}};\\
 {\displaystyle \sin^2\eta={\sin^2\beta}/{\sin^2\theta}}.
  \end{array}\right.
\end{equation}
Now, applying set (\ref{eq:new pair}) to (\ref{eq:r}), the square of the dimensionless function $\bar{r}$ becomes as
\begin{equation}\label{eq:square of r}
     \bar{r}^2=\frac{\cos^2\beta(1+\sin^2\alpha)(\sin^2\beta+\cos^2\beta\sin^2\alpha)}{{\begin{array}{l}
                                                                                               (\sin\varphi\cos\beta\sin\alpha- \cos\varphi\sin\beta)^2+\\
                                                                                               \cos^2\beta(1+\sin^2\alpha)(\cos\varphi\cos\beta\sin\alpha+\sin\varphi\sin\beta)^2
                                                                                             \end{array}}}.
\end{equation}
Then, for the dimensionless  parameter  $ \bar{z}_{\lambda}$, we have
\begin{equation}\label{eq:z_lambda}
  \bar{z}_{\lambda}=\bar{z}_B+\bar{r}\frac{\sin\varphi\cos\beta\sin\alpha- \cos\varphi\sin\beta}{\sqrt{\cos^2\beta(1+\sin^2\alpha)}}.
\end{equation}

Substituting (\ref{eq:square of r}), (\ref{eq:z_lambda}) and
\begin{equation}\label{eq:terms}
 \begin{array}{l}
  {\displaystyle t_1=\bar{x}_B+\bar{y}_B\cdot\bar{r}^2\frac{\sin^2\beta+\cos^2\beta\sin^2\alpha}{\cos^2\beta(1+\sin^2\alpha)}\times}\\
    {\displaystyle \;\;\;\;\;\;\;\;\;\;\;\;\;\;\;\;\;\;\;\;\;\;\;\;(\sin\varphi\cos\beta\sin\alpha- \cos\varphi\sin\beta)\times}\\
       {\displaystyle\;\;\;\;\;\;\;\;\;\;\;\;\;\;\;\;\;\;\;\;\;\;\;\;\;\;\;\; (\cos\varphi\cos\beta\sin\alpha+\sin\varphi\sin\beta);}\\
  {\displaystyle t_2=\bar{y}_B-\bar{x}_B\cdot\bar{r}^2\frac{\sin^2\beta+\cos^2\beta\sin^2\alpha}{\cos^2\beta(1+\sin^2\alpha)}\times}\\
    {\displaystyle \;\;\;\;\;\;\;\;\;\;\;\;\;\;\;\;\;\;\;\;\;\;\;\;(\sin\varphi\cos\beta\sin\alpha- \cos\varphi\sin\beta)\times}\\
       {\displaystyle\;\;\;\;\;\;\;\;\;\;\;\;\;\;\;\;\;\;\;\;\;\;\;\;\;\;\;\; (\cos\varphi\cos\beta\sin\alpha+\sin\varphi\sin\beta),}
 \end{array}
\end{equation}
into Eq. (\ref{eq:NEW FORMULA}), the angular misalignment of the secondary circle are defined through the pair of angle $\alpha$ and $\beta$ corresponding to the manner II.

For the case when the two circles are mutually perpendicular to each other, assuming that $\alpha=\pi/2$ then just replacing the angle $\eta$ by $\beta$ in formula (\ref{eq:Singular case}), it can be  used for calculation with new pair of angle $\alpha$ and $\beta$.

Since formula (\ref{eq:PROJECTOIN FORMULA}) is a particular case of (\ref{eq:NEW FORMULA}), when $\bar{s}=0$ and $\bar{r}=\bar{\rho}=1$. Hence, substituting $\bar{r}=1$ into (\ref{eq:z_lambda}) and using this modified equation for  formula (\ref{eq:PROJECTOIN FORMULA}), it can be used for calculation of mutual inductance between the primary circle and its projection on a tilted plane, an angular position  of which  is defined by the pair of angle $\alpha$ and $\beta$.

\section*{References}

\bibliography{References}

\begin{thebibliography}{10}
\expandafter\ifx\csname url\endcsname\relax
  \def\url#1{\texttt{#1}}\fi
\expandafter\ifx\csname urlprefix\endcsname\relax\def\urlprefix{URL }\fi
\expandafter\ifx\csname href\endcsname\relax
  \def\href#1#2{#2} \def\path#1{#1}\fi

\bibitem{Grover1944}
F.~W. Grover, The calculation of the mutual inductance of circular filaments in
  any desired positions, Proceedings of the IRE 32~(10) (1944) 620--629.
\newblock \href {http://dx.doi.org/10.1109/JRPROC.1944.233364}
  {\path{doi:10.1109/JRPROC.1944.233364}}.

\bibitem{BabicSiroisAkyelEtAl2010}
S.~Babic, F.~Sirois, C.~Akyel, C.~Girardi,
  \href{https://doi.org/10.1109/TMAG.2010.2047651}{Mutual inductance
  calculation between circular filaments arbitrarily positioned in space:
  Alternative to {G}rover's formula}, IEEE Transactions on Magnetics 46~(9)
  (2010) 3591--3600.
\newblock \href {http://dx.doi.org/10.1109/TMAG.2010.2047651}
  {\path{doi:10.1109/TMAG.2010.2047651}}.
\newline\urlprefix\url{https://doi.org/10.1109/TMAG.2010.2047651}

\bibitem{Rosa1908}
E.~B. Rosa, The self and mutual inductances of linear conductors, US Department
  of Commerce and Labor, Bureau of Standards, 1908.

\bibitem{Grover2004}
F.~W. Grover, Inductance calculations : working formulas and tables, special
  ed. prepared for instrument society of america Edition, Research Triangle
  Park, N.C. : Instrument Society of America, 1981, reprint. Originally
  published: New York : Van Nostrand, 1946. With publisher's comment.

\bibitem{Dwight1945}
H.~B. Dwight, Electrical Coils and Conductors: Their Electrical
  Characteristics, McGraw-Hill, 1945.

\bibitem{Snow1954}
C.~Snow, Formulas for computing capacitance and inductance, Vol. 544, US Govt.
  Print. Off., 1954.

\bibitem{Zeitlin1950}
L.~A. Zeitlin, Induktivnosti provodov i konturov (Inductances of wires and
  loops), Gosenergoizdat, Leningrad - Moskva, 1950.

\bibitem{Kalantarov1986}
P.~L. Kalantarov, L.~A. Zeitlin, Raschet induktivnostey (Calculation of
  Inductances), 3rd Edition, Energoatomizdat, Leningrad, 1986.

\bibitem{KamonTsukWhite1994}
M.~Kamon, M.~J. Tsuk, J.~K. White, Fasthenry: a multipole-accelerated 3-{D}
  inductance extraction program, IEEE Transactions on Microwave Theory and
  Techniques 42~(9) (1994) 1750--1758.
\newblock \href {http://dx.doi.org/10.1109/22.310584}
  {\path{doi:10.1109/22.310584}}.

\bibitem{OkressWroughtonComenetzEtAl1952}
E.~Okress, D.~Wroughton, G.~Comenetz, P.~Brace, J.~Kelly, Electromagnetic
  levitation of solid and molten metals, Journal of Applied Physics 23~(5)
  (1952) 545--552.

\bibitem{Urman1997}
Y.~M. Urman, \href{https://doi.org/10.1134/1.1258645}{Theory for the
  calculation of the force characteristics of an electromagnetic suspension of
  a superconducting body}, Technical Physics 42~(1) (1997) 1--6.
\newblock \href {http://dx.doi.org/10.1134/1.1258645}
  {\path{doi:10.1134/1.1258645}}.
\newline\urlprefix\url{https://doi.org/10.1134/1.1258645}

\bibitem{Urman1997a}
Y.~M. Urman, \href{https://doi.org/10.1134/1.1258654}{Calculation of the force
  characteristics of a multi-coil suspension of a superconducting sphere},
  Technical Physics 42~(1) (1997) 7--13.
\newblock \href {http://dx.doi.org/10.1134/1.1258654}
  {\path{doi:10.1134/1.1258654}}.
\newline\urlprefix\url{https://doi.org/10.1134/1.1258654}

\bibitem{Coffey2001}
M.~W. Coffey, Mutual inductance of superconducting thin films, Journal of
  Applied Physics 89~(10) (2001) 5570--5577.

\bibitem{Urman2014}
Y.~Urman, S.~Kuznetsov, Translational transformations of tensor solutions of
  the helmholtz equation and their application to describe interactions in
  force fields of various physical nature, Quarterly of Applied Mathematics
  72~(1) (2014) 1--20.

\bibitem{JowGhovanloo2007}
U.-M. Jow, M.~Ghovanloo, Design and optimization of printed spiral coils for
  efficient transcutaneous inductive power transmission, IEEE Transactions on
  biomedical circuits and systems 1~(3) (2007) 193--202.

\bibitem{SuLiuHui2009}
Y.~P. Su, X.~Liu, S.~Y.~R. Hui, Mutual inductance calculation of movable planar
  coils on parallel surfaces, IEEE Transactions on Power Electronics 24~(4)
  (2009) 1115--1123.
\newblock \href {http://dx.doi.org/10.1109/TPEL.2008.2009757}
  {\path{doi:10.1109/TPEL.2008.2009757}}.

\bibitem{ChuAvestruz2017}
S.~Y. Chu, A.~T. Avestruz, Transfer-power measurement: A non-contact method for
  fair and accurate metering of wireless power transfer in electric vehicles,
  in: 2017 IEEE 18th Workshop on Control and Modeling for Power Electronics
  (COMPEL), 2017, pp. 1--8.
\newblock \href {http://dx.doi.org/10.1109/COMPEL.2017.8013344}
  {\path{doi:10.1109/COMPEL.2017.8013344}}.

\bibitem{ShiriShoulaie2009}
A.~Shiri, A.~Shoulaie, A new methodology for magnetic force calculations
  between planar spiral coils, Progress In Electromagnetics Research 95 (2009)
  39--57.

\bibitem{RavaudLemarquandLemarquand2009}
R.~Ravaud, G.~Lemarquand, V.~Lemarquand, Force and stiffness of passive
  magnetic bearings using permanent magnets. part 1: Axial magnetization, IEEE
  transactions on magnetics 45~(7) (2009) 2996.

\bibitem{Obata2013}
S.~Obata, A muscle motion solenoid actuator, Electrical Engineering in Japan
  184~(2) (2013) 10--19.

\bibitem{ShalatiPoletkinKorvinkEtAl2018}
R.~Shalati, K.~V. Poletkin, J.~G. Korvink, V.~Badilita,
  \href{http://stacks.iop.org/1742-6596/1052/i=1/a=012047}{Novel concept of a
  series linear electromagnetic array artificial muscle}, Journal of Physics:
  Conference Series 1052~(1) (2018) 012047.
\newline\urlprefix\url{http://stacks.iop.org/1742-6596/1052/i=1/a=012047}

\bibitem{Poletkin2013}
K.~Poletkin, A.~I. Chernomorsky, C.~Shearwood, U.~Wallrabe,
  \href{http://dx.doi.org/10.1115/IMECE2013-66010}{An analytical model of
  micromachined electromagnetic inductive contactless suspension.}, in: the
  ASME 2013 International Mechanical Engineering Congress \& Exposition, ASME,
  San Diego, California, USA, 2013, pp. V010T11A072--V010T11A072.
\newblock \href {http://dx.doi.org/10.1115/IMECE2013-66010}
  {\path{doi:10.1115/IMECE2013-66010}}.
\newline\urlprefix\url{http://dx.doi.org/10.1115/IMECE2013-66010}

\bibitem{Poletkin2014a}
K.~Poletkin, A.~Chernomorsky, C.~Shearwood, U.~Wallrabe,
  \href{http://authors.elsevier.com/sd/article/S0020740314000897}{A qualitative
  analysis of designs of micromachined electromagnetic inductive contactless
  suspension}, International Journal of Mechanical Sciences 82 (2014) 110--121.
\newblock \href {http://dx.doi.org/10.1016/j.ijmecsci.2014.03.013}
  {\path{doi:10.1016/j.ijmecsci.2014.03.013}}.
\newline\urlprefix\url{http://authors.elsevier.com/sd/article/S0020740314000897}

\bibitem{Lu2014}
Z.~Lu, K.~Poletkin, B.~den Hartogh, U.~Wallrabe, V.~Badilita,
  \href{http://dx.doi.org/10.1016/j.sna.2014.09.017}{{3D} micro-machined
  inductive contactless suspension: Testing and modeling}, Sensors and
  Actuators A Physical 220 (2014) 134--143.
\newblock \href {http://dx.doi.org/10.1016/j.sna.2014.09.017}
  {\path{doi:10.1016/j.sna.2014.09.017}}.
\newline\urlprefix\url{http://dx.doi.org/10.1016/j.sna.2014.09.017}

\bibitem{PoletkinLuWallrabeEtAl2017b}
K.~Poletkin, Z.~Lu, U.~Wallrabe, J.~Korvink, V.~Badilita,
  \href{http://www.sciencedirect.com/science/article/pii/S0020740316306555}{Stable
  dynamics of micro-machined inductive contactless suspensions}, International
  Journal of Mechanical Sciences 131-132 (2017) 753 -- 766.
\newblock \href
  {http://dx.doi.org/https://doi.org/10.1016/j.ijmecsci.2017.08.016}
  {\path{doi:https://doi.org/10.1016/j.ijmecsci.2017.08.016}}.
\newline\urlprefix\url{http://www.sciencedirect.com/science/article/pii/S0020740316306555}

\bibitem{Poletkin2012}
K.~V. Poletkin, A.~I. Chernomorsky, C.~Shearwood, A proposal for micromachined
  accelerometer, base on a contactless suspension with zero spring constant,
  IEEE Sensors J. 12~(07) (2012) 2407--2413.
\newblock \href {http://dx.doi.org/10.1109/JSEN.2012.2188831}
  {\path{doi:10.1109/JSEN.2012.2188831}}.

\bibitem{PoletkinShalatiKorvinkEtAl2018}
K.~V. Poletkin, R.~Shalati, J.~G. Korvink, V.~Badilita,
  \href{http://stacks.iop.org/1742-6596/1052/i=1/a=012035}{Pull-in actuation in
  hybrid micro-machined contactless suspension}, Journal of Physics: Conference
  Series 1052~(1) (2018) 012035.
\newline\urlprefix\url{http://stacks.iop.org/1742-6596/1052/i=1/a=012035}

\bibitem{PoletkinKorvink2018}
K.~V. Poletkin, J.~G. Korvink, Modeling a pull-in instability in micro-machined
  hybrid contactless suspension 7~(1) (2018) 11.

\bibitem{TheodoulidisDitchburn2007}
T.~Theodoulidis, R.~J. Ditchburn, Mutual impedance of cylindrical coils at an
  arbitrary position and orientation above a planar conductor, IEEE
  Transactions on Magnetics 43~(8) (2007) 3368--3370.
\newblock \href {http://dx.doi.org/10.1109/TMAG.2007.894559}
  {\path{doi:10.1109/TMAG.2007.894559}}.

\bibitem{SawanHashemiSehilEtAl2009}
M.~Sawan, S.~Hashemi, M.~Sehil, F.~Awwad, M.~Hajj-Hassan, A.~Khouas,
  \href{https://doi.org/10.1007/s10544-009-9323-7}{Multicoils-based inductive
  links dedicated to power up implantable medical devices: modeling, design and
  experimental results}, Biomedical Microdevices 11~(5) (2009) 1059.
\newblock \href {http://dx.doi.org/10.1007/s10544-009-9323-7}
  {\path{doi:10.1007/s10544-009-9323-7}}.
\newline\urlprefix\url{https://doi.org/10.1007/s10544-009-9323-7}

\bibitem{KuznetsovGuest2017}
S.~Kuznetsov, J.~K. Guest,
  \href{http://www.sciencedirect.com/science/article/pii/S0304885316319515}{Topology
  optimization of magnetic source distributions for diamagnetic and
  superconducting levitation}, Journal of Magnetism and Magnetic Materials 438
  (2017) 60 -- 69.
\newblock \href {http://dx.doi.org/https://doi.org/10.1016/j.jmmm.2017.04.052}
  {\path{doi:https://doi.org/10.1016/j.jmmm.2017.04.052}}.
\newline\urlprefix\url{http://www.sciencedirect.com/science/article/pii/S0304885316319515}

\bibitem{D.I.B.2002}
D.~Hoult, B.~Tomanek,
  \href{https://onlinelibrary.wiley.com/doi/abs/10.1002/cmr.10047}{Use of
  mutually inductive coupling in probe design}, Concepts in Magnetic Resonance
  15~(4) (2002) 262--285.
\newblock \href
  {http://arxiv.org/abs/https://onlinelibrary.wiley.com/doi/pdf/10.1002/cmr.10047}
  {\path{arXiv:https://onlinelibrary.wiley.com/doi/pdf/10.1002/cmr.10047}},
  \href {http://dx.doi.org/10.1002/cmr.10047} {\path{doi:10.1002/cmr.10047}}.
\newline\urlprefix\url{https://onlinelibrary.wiley.com/doi/abs/10.1002/cmr.10047}

\bibitem{SpenglerWhileMeissnerEtAl2017}
N.~Spengler, P.~T. While, M.~V. Meissner, U.~Wallrabe, J.~G. Korvink,
  \href{https://doi.org/10.1371/journal.pone.0182779}{Magnetic lenz lenses
  improve the limit-of-detection in nuclear magnetic resonance}, PLOS ONE
  12~(8) (2017) 1--17.
\newblock \href {http://dx.doi.org/10.1371/journal.pone.0182779}
  {\path{doi:10.1371/journal.pone.0182779}}.
\newline\urlprefix\url{https://doi.org/10.1371/journal.pone.0182779}

\bibitem{AngelisPaskuAngelisEtAl2015}
G.~D. Angelis, V.~Pasku, A.~D. Angelis, M.~Dionigi, M.~Mongiardo, A.~Moschitta,
  P.~Carbone, An indoor ac magnetic positioning system, IEEE Transactions on
  Instrumentation and Measurement 64~(5) (2015) 1267--1275.
\newblock \href {http://dx.doi.org/10.1109/TIM.2014.2381353}
  {\path{doi:10.1109/TIM.2014.2381353}}.

\bibitem{WuJeonMoonEtAl2016}
F.~Wu, J.~Jeon, S.~K. Moon, H.~J. Choi, H.~Son, Voice coil navigation sensor
  for flexible silicone intubation, IEEE/ASME Transactions on Mechatronics
  21~(2) (2016) 851--859.
\newblock \href {http://dx.doi.org/10.1109/TMECH.2015.2476836}
  {\path{doi:10.1109/TMECH.2015.2476836}}.

\bibitem{Gulbahar2017}
B.~Gulbahar, A communication theoretical analysis of multiple-access channel
  capacity in magneto-inductive wireless networks, IEEE Transactions on
  Communications 65~(6) (2017) 2594--2607.
\newblock \href {http://dx.doi.org/10.1109/TCOMM.2017.2669995}
  {\path{doi:10.1109/TCOMM.2017.2669995}}.

\bibitem{Maxwell1954}
J.~C. Maxwell, A Treatise on Electricity and Magnetism, 3rd Edition, Vol.~2,
  Dover Publications Inc., 1954.

\bibitem{ButterworthM.Sc.1916}
S.~Butterworth, \href{https://doi.org/10.1080/14786440508635521}{{LIII. O}n the
  coefficients of mutual induction of eccentric coils}, The London, Edinburgh,
  and Dublin Philosophical Magazine and Journal of Science 31~(185) (1916)
  443--454.
\newblock \href
  {http://arxiv.org/abs/https://doi.org/10.1080/14786440508635521}
  {\path{arXiv:https://doi.org/10.1080/14786440508635521}}, \href
  {http://dx.doi.org/10.1080/14786440508635521}
  {\path{doi:10.1080/14786440508635521}}.
\newline\urlprefix\url{https://doi.org/10.1080/14786440508635521}

\bibitem{Snow1928}
C.~Snow, Mutual inductance of any two circles, Bur. Stand. J. Res 1 (1928)
  531--542.

\bibitem{SpiegelLipschutzLiu2009}
M.~R. Spiegel, S.~Lipschutz, J.~Liu, Schaum's outline of mathematical handbook
  of formulas and tables, 3rd Edition, McGraw-Hill New York, 2009.

\end{thebibliography}

\end{document}